\newcommand{\degree}{$^{\circ}$}

\documentclass{pasa}%

\title[A Census of Southern Pulsars at 185\,MHz]{A Census of Southern Pulsars at 185\,MHz}
\author[Mengyao Xue et al.]{Mengyao Xue$^{1,2}$, N. D. R. Bhat$^{1,2}$, S. E. Tremblay$^{1,2}$, S. M. Ord$^{3}$, C. Sobey$^{1,4}$, N. A. Swainston$^{1,2}$, D. L. Kaplan$^{5}$, Simon Johnston$^{3}$, B. W. Meyers$^{1,2,3}$, \and S. J. McSweeney$^{1,2}$\\
\affil{$^{1}$International Centre for Radio Astronomy Research (ICRAR), Curtin University, Bentley, WA 6102, Australia}%
\affil{$^{2}$ARC Centre of Excellence for All-sky Astrophysics (CAASTRO)}%
\affil{$^{3}$CSIRO Astronomy and Space Science, Australia Telescope National Facility, PO Box 76, Epping, NSW 1710, Australia}%
\affil{$^{4}$CSIRO Astronomy and Space Science, 26 Dick Perry Avenue, Kensington, WA 6151, Australia}%
\affil{$^{5}$Department of Physics, University of Wisconsin-Milwaukee, Milwaukee, WI 53201, USA}}%

\jid{PASA}
\doi{10.1017/pas.\the\year.xxx}
\jyear{\the\year}

\usepackage[authoryear]{natbib}
\bibpunct{(}{)}{;}{a}{}{,}
\usepackage{aas_macros}
\usepackage{hyperref}
\usepackage{graphicx}
\usepackage{subfigure}
\usepackage{caption}
\usepackage{longtable}
\usepackage{threeparttablex}
\usepackage{pdflscape}
\usepackage{array}

\hypersetup{colorlinks,citecolor=blue,linkcolor=blue,urlcolor=blue}

\begin{document}%
\begin{abstract}
The Murchison Widefield Array (MWA), and its recently-developed Voltage Capture System (VCS), facilitates extending the low-frequency range of pulsar observations at high-time and -frequency resolution in the Southern Hemisphere, providing further information about pulsars and the ISM. We present the results of an initial time-resolved census of known pulsars using the MWA. 
To significantly reduce the processing load, we incoherently sum the detected powers from the 128 MWA tiles, which yields $\sim10\%$ of the attainable sensitivity of the coherent sum. This preserves the large field-of-view ($\sim$450\,deg$^2$ at 185\,MHz), allowing multiple pulsars to be observed simultaneously. We developed a WIde-field Pulsar Pipeline (WIPP) that processes the data from each observation and automatically folds every known pulsar located within the beam.
We have detected 50 pulsars to date, 6 of which are millisecond pulsars. This is consistent with our expectation, given the telescope sensitivity and the sky coverage of the processed data ($\sim$17,000\,deg$^2$). For ten pulsars, we present the lowest-frequency detections published. For a subset of the pulsars, we present multi-frequency pulse profiles by combining our data with published profiles from other telescopes.
Since the MWA is a low-frequency precursor to the Square Kilometre Array (SKA), we use our census results to forecast that a survey using Phase 1 of SKA-Low (SKA1-Low) can potentially detect around 9400 pulsars.
\end{abstract}
\maketitle%

\section{INTRODUCTION} \label{sec:intro}

Pulsars were discovered through observing their pulsed emission at a very low radio frequency of 81.5\,MHz \citep[][]{Hewish1968}.  Although low-frequency observations ($<$300\,MHz) played a considerable role in early pulsar science \citep[e.g.][]{Taylor1977,Taylor1986}, the vast majority of the 2536 pulsars now catalogued\footnote{Version 1.54 of the pulsar catalogue retrieved from www.atnf.csiro.au/people/pulsar/psrcat}\citep{PSRCAT2005} were discovered and studied at higher frequencies, between 300\,MHz and a few GHz. This is due to a compromise between three frequency-dependent effects (1) the greater broadening of pulsed signals towards lower frequencies due to interstellar medium (ISM) propagation effects, particularly multipath scattering, which is a strong function of the observing frequency ($\propto\nu^{\rm{-4}}$, e.g. \citealt{Bhat2004}), (2) the increase in telescope system temperature towards lower frequencies due to the diffuse Galactic continuum emission ($\propto\nu^{\rm{-2.6}}$; \citealt{Lawson:1987}), and (3) the decrease in flux densities towards higher frequencies due to pulsars' steep intrinsic spectral indices ($\propto\nu^{\rm{-1.4}}$, on average; \citealt{Bates2013}).

Pulsars are once again being routinely observed and studied at low frequencies due to advances in instrumentation and computing. This includes recently upgraded or constructed telescopes operating below 300\,MHz and their associated computing facilities: the Giant Metrewave Radio Telescope {\citep[GMRT;][]{GMRT1991,Roy2010ExA}, the Long Wavelength Array {\citep[LWA;][]{Taylor2012,Stovall2015}}, the Low-Frequency Array {\citep[LOFAR;][]{LOFARproject2013,Stappers2011}}, and the Murchison Widefield Array {\citep[MWA;][]{Tingay2013,Tremblay2015}}. Besides the MWA, all of these facilities are located in the Northern Hemisphere. Consequently, 22\% of the known pulsar population can only be observed at low frequencies using the MWA ($-90$\degree$\leqslant\delta\lesssim-50$\degree). Moreover, the sensitivity of aperture array telescopes decreases rapidly for lower elevation pointings \citep[$\lesssim$30\degree;][]{Noutsos:2015,Stovall2015}, corresponding to declinations below approximately $-10$\degree~for LOFAR and $-30$\degree~for the LWA and above approximately $+30$\degree~for the MWA. 
However, these telescopes are also situated over a range of longitudes, and there is a significant overlap in the area of observable sky, as shown in Figure~\ref{fig:01}. This is particularly useful for monitoring observations and verifying results \citep[e.g.][]{Hermsen2013,Dolch:2014,Mereghetti2016}.

\begin{figure*}[ht]
\begin{center}
\includegraphics[scale=0.13, angle=0]{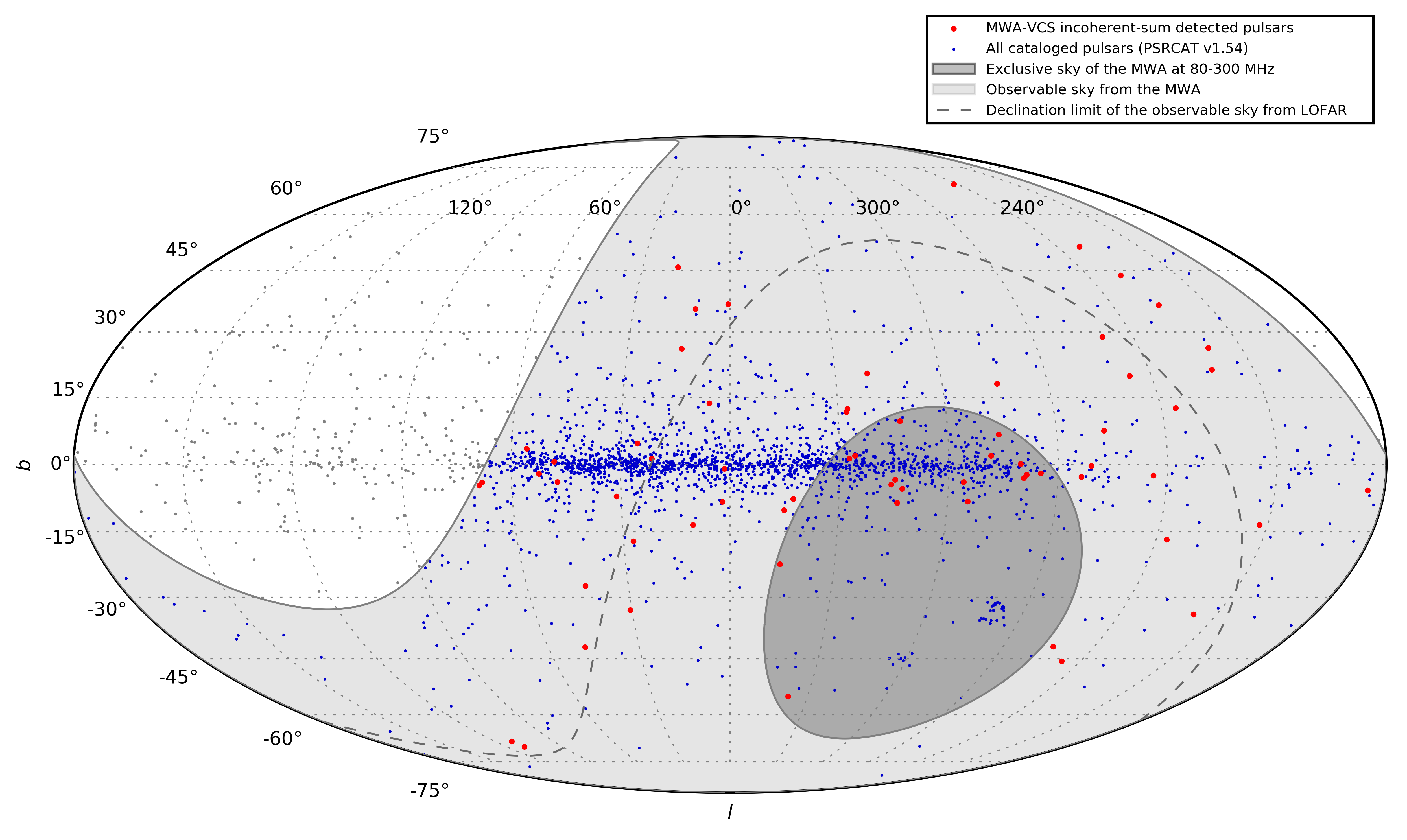}
\caption{The locations of catalogued pulsars in Galactic coordinates (blue points; using version 1.54 of the pulsar catalogue). The observable sky from the MWA telescope ($\delta\lesssim+30$\degree) and the area where the MWA can exclusively observe pulsars below 300\,MHz ($\delta\lesssim-50$\degree) are indicated by the light and dark shaded regions, respectively. The grey dashed line indicates the declination limit of the observable sky from LOFAR ($\delta\gtrsim-10$\degree). We note that for aperture arrays like the MWA and LOFAR, observations can be made at any zenith angle, the sensitivity falls off with the zenith angle, and is significantly reduced when pointed to elevations $\lesssim 30$\degree~\citep{Noutsos:2015}. In this figure, we use hard limits purely for illustration. The positions of the pulsars detected at 185\,MHz in this work, using the MWA in incoherent-sum mode, are also shown (red circles).}\label{fig:01}
\end{center}
\end{figure*}

Low-frequency observations of pulsars provide insights into the physics of the radio emission mechanism and the ISM \citep[e.g.][]{Stappers2011}, especially in a multi-frequency context. For example, the widths of integrated pulse profiles generally widen with decreasing observing frequencies \citep[e.g.][]{Johnston2008}. This behaviour is often explained by the radius-to-frequency mapping (RFM) model, whereby lower-frequency emission arises from higher altitudes in the pulsar magnetosphere \citep[e.g.][]{Cordes1978, Mitra&Rankin2002}. Therefore, we assume that pulse profiles observed at a range of frequencies allow us to trace the open dipolar field line region of the pulsar magnetosphere, which increases in size towards higher altitudes. The simplest pulse profiles often show the clearest examples of RFM, while multi-component profiles often show more complex behaviour \citep{Johnston2008}. Furthermore, pulse profiles are often observed to show an increasingly rapid evolution towards lower frequencies \citep[e.g.][]{Bhat2014,LOFAR100profile2016}. This is possibly due to the frequency-dependence of the emission beam opening angle \citep{Thorsett1991,Xilouris1996}, which also depends on the pulsar spin period and spin-down \citep{Kijak1998a,Kijak2003}. Therefore, comparing low-frequency observations to those at higher frequencies for a representative sample of pulsars can enable us to understand the beam geometry better and provide further insights into the enigmatic radio emission mechanism from pulsar magnetospheres.

Low-frequency pulsar observations also provide precise measurements of properties of the ionised ISM, including electron densities \citep[e.g.][]{Hassall:2012,Bilous:2016} and magnetic fields \citep[e.g.][]{Noutsos:2015,Howard:2016}, due to the steep power-law dependence on frequency of the propagation effects. For example, \cite{Stovall2015} were able to study the ISM propagation effects using 44 pulsars detected at frequencies less than 100\,MHz with the LWA.

A major motivation for the Square Kilometre Array (SKA) is the science facilitated using pulsar observations. The MWA is the low-frequency precursor telescope to the SKA. Therefore, pulsar observations using the same observing environment and frequencies are necessary to prepare for pulsar science with the SKA-Low. 

In this work, we present an initial low-frequency census of known (catalogued) pulsars in the Southern Hemisphere. The descriptions of our MWA observations and data processing methods are outlined in Section 2. In Section 3 we provide a summary of the 50 pulsars detected to date. We discuss our results in Section 4, including the investigation of the population of pulsars observable using SKA-Low, and summarise our work in Section 5.

\section{OBSERVATIONS AND ANALYSIS}

\subsection{The MWA and the Voltage Capture System} \label{sec:vcs}

The MWA is located at the Murchison Radio-Astronomy Observatory (MRO) in Western Australia. The remote, radio-quiet environment minimises radio frequency interference (RFI). The MWA consists of an array of 128 tiles, distributed with baselines up to $\sim$3\,km, where each tile comprises 16 dual-polarisation dipole antennas arranged in a regular $4\times 4$ grid. Since the array has no moving parts, tile beams are formed by electronic manipulation of the dipole signals in an analogue beamformer \citep{Tingay2013}. A bandwidth of 30.72\,MHz can be flexibly recorded over a frequency range of 80--300\,MHz. 

Although the MWA was originally designed as an imaging interferometer, the Voltage Capture System {\citep[VCS;][]{Tremblay2015}} has extended the capabilities of this telescope, allowing it to record high-time and frequency resolution voltage data. VCS observations are channelised to a frequency resolution of 10\,kHz and a time resolution of 100\,$\mu$s, allowing the MWA to provide phase-resolved observations of pulsars \citep[e.g.][]{Bhat2014}. Commissioning of the MWA-VCS was completed in 2014. 

VCS data recorded from each MWA tile can be reduced in different ways, providing its user with maximum flexibility. The least compute-intensive method is to perform an incoherent sum: voltages from each tile are multiplied by their complex conjugate to form the power, and then summed together. This incoherent sum preserves the full single tile field-of-view (FoV; $\sim$450\,deg$^2$ at 185\,MHz) and increases the sensitivity by a factor of $\sqrt{N}$ over a single tile theoretically, where $N$ is the number of tiles summed. The VCS raw data can also be processed to generate a coherent sum of the tile signals by applying a phase rotation to each voltage stream and then summing to increase the sensitivity over a single tile by a factor of $N$ theoretically, and reducing the FoV to $\sim$2 arcmin in diameter. Forming a single coherent tied-array beam requires 2-3 times more compute time, in addition to manually-generated calibration solutions (e.g. \citealt{Bhat2016}). 
We therefore conducted an initial pulsar census using the incoherently summed data. Pulsars detected using the coherently beamformed data will be reported in a future publication.

\subsection{Observations} \label{sec:observations}

Approximately 84 hours of observations were made at a frequency of 185\,MHz during the first two years of MWA-VCS operations\footnote{This frequency was chosen because the beam is comparatively well-understood and the band is nearly free from RFI}. These data are from 104 observing sessions, ranging from 5\,minutes to 1.5\,hours in duration, and were collected for a variety of scientific projects, e.g. pulsar emission studies, millisecond pulsar monitoring, and searches for Fast Radio Bursts (e.g. \citealt{Thornton2013}). These observations are drift scans, i.e. the elevation and azimuth were fixed throughout the pointing instead of tracking a particular source. Each observation is identified by the GPS start time (`OBS ID' hereafter).
The VCS system records data at a rate of 28\,TB/hr, and the total size of the entire data set is approximately 2.6\,PB\footnote{1\,petabyte (PB) = 1024\,terabytes (TB)}.
These raw data are archived at the Pawsey Supercomputing Centre.

We processed 37 hours of VCS data from 46 individual observing sessions that cover the sky efficiently, i.e. we have excluded duplicate sky pointings. These observations amount to $\sim$17,000 deg$^2$ of the sky (albeit with varying sensitivity), which corresponds to $\sim$55\% of the whole $\delta \lesssim 30$\degree~sky (see Figure~\ref{fig:02}).

\begin{figure*}[!ht]
\begin{center}
\includegraphics[scale=0.32, angle=0]{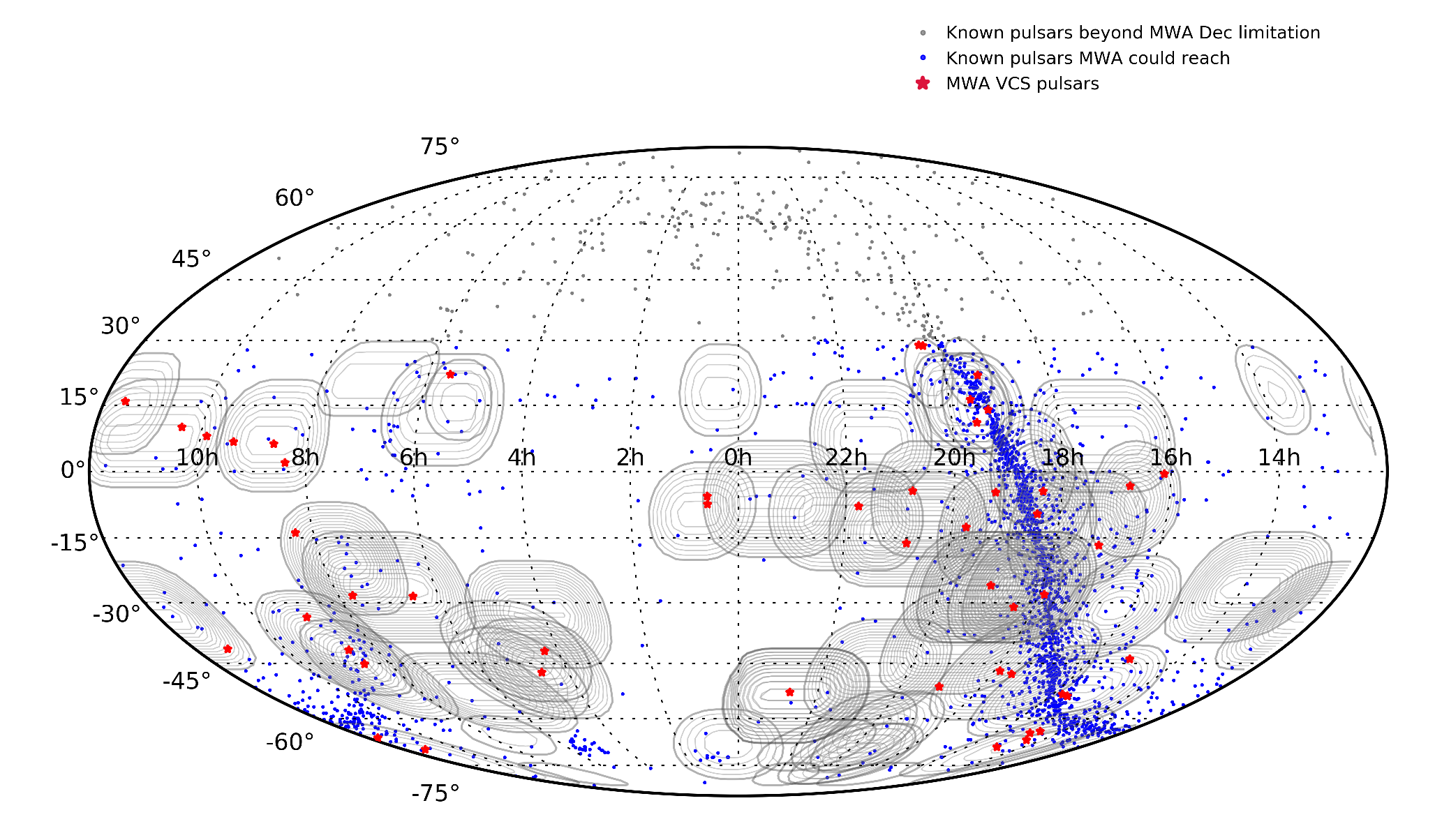}
\caption{The sky coverage of the MWA-VCS observations processed in this work. Grey contour lines represent the beam pattern (for beam powers greater than 25\% of that at zenith); blue points indicate all known pulsars (catalogue v1.54); red stars show the pulsars detected in this work.}\label{fig:02}
\end{center}
\end{figure*}

\subsection{Pre-processing}

As described in \cite{Tremblay2015}, the channelised raw voltage data are recorded in 32 files, each being a fine polyphase filter bank (PFB) output per tile per polarisation per second. We generate an incoherent beam (as described in Section\,\ref{sec:vcs}) for each second of the observation, and write out every 200-seconds of data to a file in the PSRFITS format \citep{Hotan2004}, retaining a frequency resolution of 10\,kHz and a time resolution of 100\,$\mu$s. Therefore, each observation typically consists of multiple PSRFITS files. These PSRFITS files are then processed by the WIde-field Pulsar Pipeline (WIPP), described in Section \ref{sec:pipeline}.

\subsection{Processing pipeline} \label{sec:pipeline}

The incoherently-summed MWA data preserves the large FoV and, therefore, multiple pulsars may be located within the tile beam in each observation. We developed a WIde-field Pulsar Pipeline (WIPP) that automatically identifies and folds all known pulsars positioned within the beam out to an established threshold.

\subsubsection{Sample Selection}

For each observation, the WIPP automatically generates a list of pulsars to fold using the pulsar catalogue (v1.54) and information about the telescope pointing direction (taking into account that during drift scans pulsars may enter and leave the beam), the beam model, and the dispersion measures (DMs) of the pulsars.
Since each MWA tile is a regularly distributed aperture array, the resulting beam pattern is complex, changing as a function of azimuth, elevation, and frequency \citep{MWAbeam2015}. An example is shown in Figure~\ref{fig:03}.
We used the MWA beam model to identify known pulsars located within the beam out to 25\% of the beam power towards zenith, at any time during the observations, and include even those located within a grating lobe in some cases. 

According to the NE2001 Galactic electron density model \citep{NE2001}, the scattering time associated with lines-of-sight towards the Galactic plane with a DM greater than 300\,cm$^{-3}$\,pc is typically more than 3 seconds. Since we are unlikely to detect pulsars with such long scattering times at 185\,MHz, we limited our selection of pulsars to those with a DM of less than 300\,cm$^{-3}$\,pc. This excluded a substantial fraction ($\sim$40\%) of the pulsars located near the Galactic plane. Moreover, we did not make any assumptions regarding the pulsars' expected flux densities and included all pulsars, whether or not they have a catalogued flux density measurement. Typically, a source list of $\sim$10--200 pulsars was generated per observation, with larger numbers for pointings near the Galactic plane.

\begin{figure*}[ht]
\begin{center}
\includegraphics[scale=0.3, angle=0]{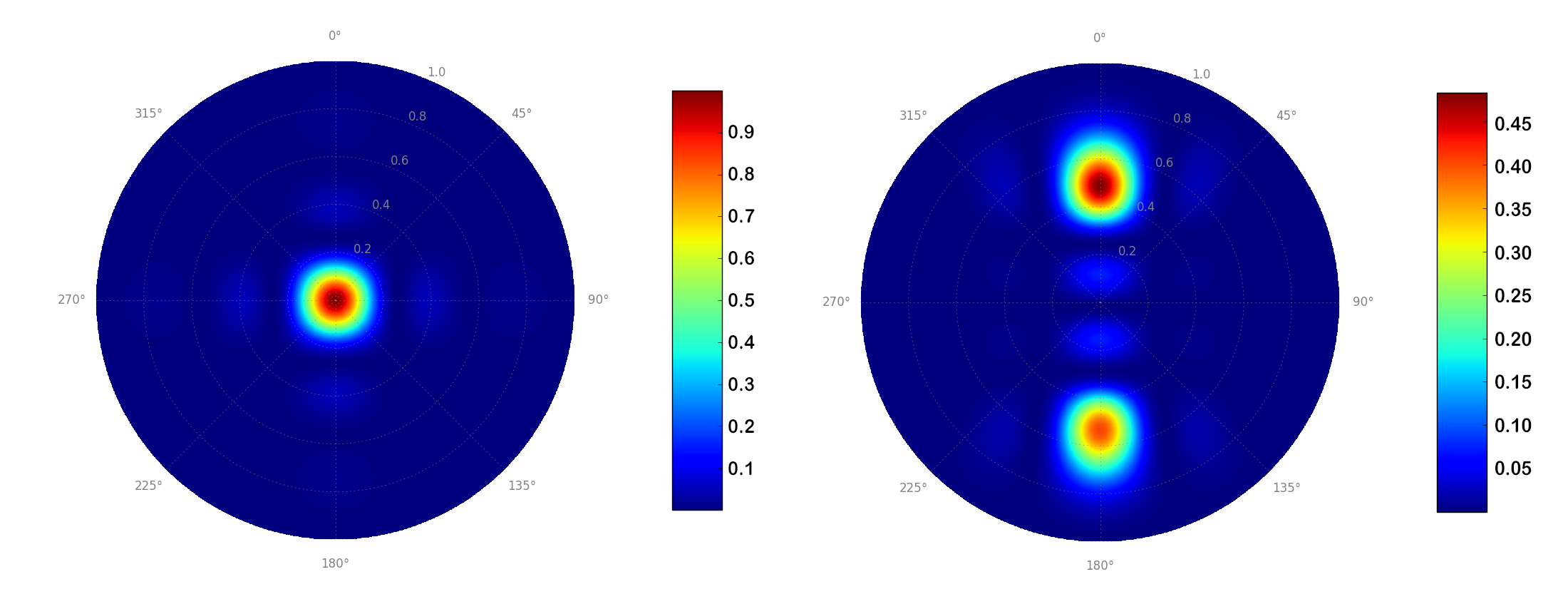}
\caption{The MWA tile beam model at 185\,MHz, indicating the power relative to pointing at zenith, see colour bars for scale. Left: an example of a pointing towards the zenith (observation ID 1088850560). Right: an example of the main beam and a grating lobe (upper and lower part of the figure, respectively; observation ID 1140972392; 0\degree~azimuth, 45\degree~elevation).}\label{fig:03}
\end{center}
\end{figure*}

\subsubsection{Pulsar Folding}

The WIPP attempts to detect every pulsar in the source list generated for each observation.
The pipeline uses the \texttt{prepfold} function of the PRESTO \footnote{www.cv.nrao.edu/~sransom/presto} software package \citep{Ransom2001} to dedisperse and fold the data using ephemerides from the pulsar catalogue.

We folded the data twice: with and without searching in DM, period, and period-derivative space. This identified changes in the ephemeris values and prevented the algorithm from locking on to the radio-frequency interference (RFI). We also applied an RFI mask output obtained from \texttt{rfifind} in PRESTO, but this did not significantly alter the results  which is likely due to the pristine RFI environment \citep{Offringa2015,Sokolowski2017}. For pulsars with rotational periods $<$6.4\,ms, we retained the 100\,$\mu$s time resolution of the data. Otherwise, 64 or 100 pulse phase bins were used for the initial detections. A flow chart of the processing procedure is shown in Figure\,\ref{fig:04}. 
After automatically folding the known pulsars in the source lists, each PRESTO output was inspected by eye to identify those successfully detected.

\begin{figure}[ht]
\begin{center}
\includegraphics[scale=0.6, angle=0]{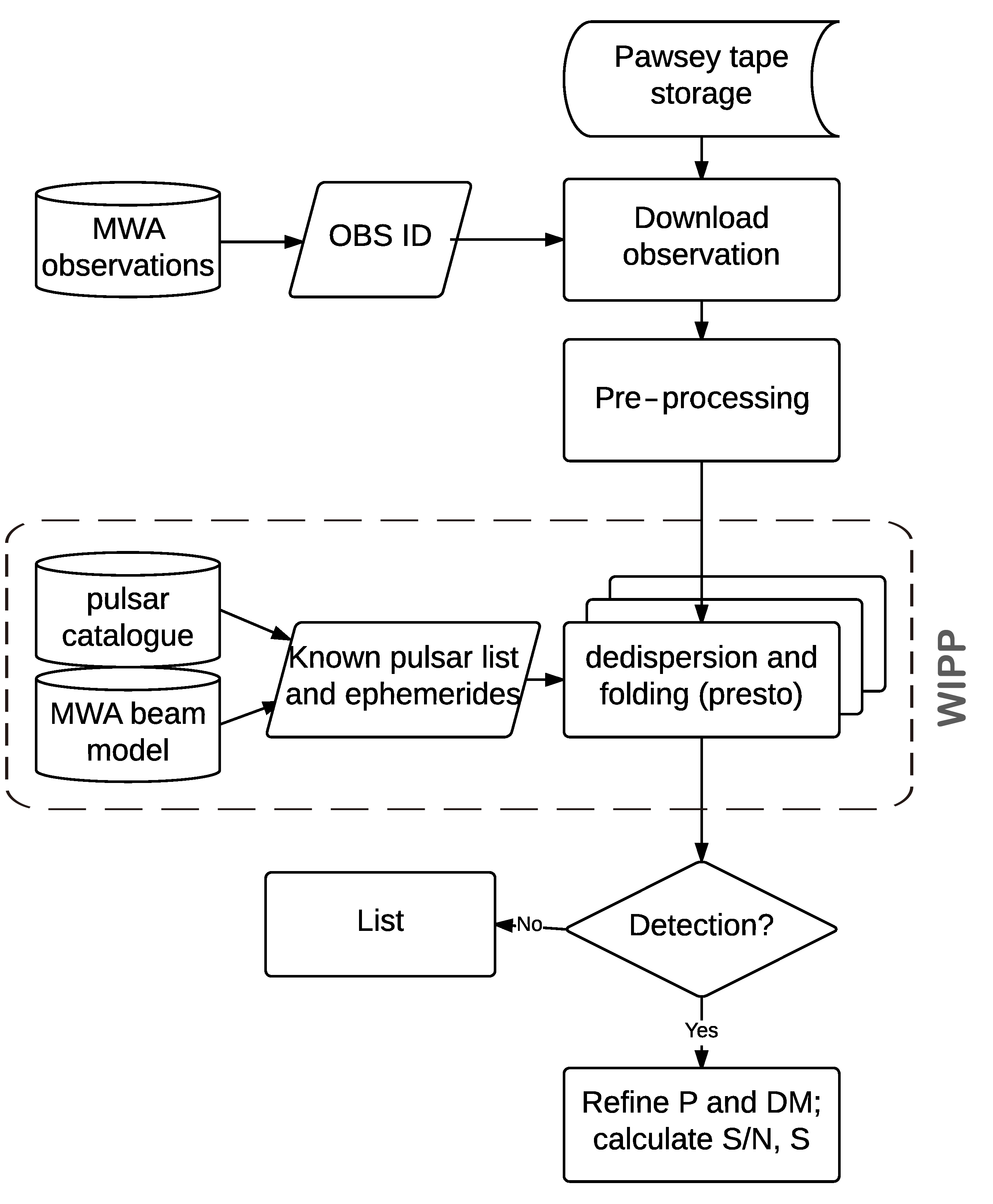}
\caption{The data processing flow: from downloading MWA-VCS archival data (from the Pawsey Supercomputing Centre) to detecting catalogued pulsars. Automatic processing procedures included in the WIPP are shown within the grey dashed line. The WIPP processes multiple pulsars in each observation. This is different from the type of processing generally employed for most traditional pulsar-capable telescopes, where there is typically only one pulsar per pointing.
}\label{fig:04}
\end{center}
\end{figure}

\subsubsection{Calculation of Flux Densities} \label{sec:flux_cal}

For each pulsar detected, we determined the peak signal-to-noise ratio, $(\text{S/N})_\text{peak}$,  
and calculated the flux density using the MWA system performance parameters, including the system temperature, $T_\text{sys}$, and the telescope gain, $G$.
The mean flux density, $S_\text{mean}$, was calculated using the radiometer equation, as applied to pulsar observations \citep[e.g][]{Lorimer2005handbook}: 
\begin{equation}\label{eq:Flux}
S_\text{mean}=\frac{(\text{S/N})_\text{peak}T_\text{sys}}{G\sqrt{n_\text{p}t_\text{int}\Delta f}} \sqrt{\frac{W}{P-W}},
\end{equation}
where $t_\text{int}$ is the integration time, $\Delta f=30.72$\,MHz is the observation bandwidth, and $n_\text{p}=2$ is the number of polarisations summed.
The quantity $P$ indicates the pulse period and $W$ represents the equivalent pulse width. 

Assuming 100\% antenna efficiency, the system temperature includes the contributions from the antenna temperature and the receiver temperature: $T_\text{sys}=T_\text{ant} + T_\text{rec}$.
Measurements of $T_\text{rec}$ for the MWA are available at different frequencies in 1.28\,MHz steps \citep[e.g. 23\,K at 184.96\,MHz and 50\,K at 87.68\,MHz;][]{Prabu2015}.
The antenna temperature, $T_\text{ant}$, for a given pointing direction in azimuth, $\theta$, and elevation, $\phi$, was determined by integrating the sky temperature $T_\text{sky}(\theta,\phi)$ over the modelled MWA beam. 
An all-sky survey at 408\,MHz \citep{Haslam1981,Haslam1982}, scaled to our observing frequency, was used to estimate $T_\text{sky}$.
For our observations, the $T_\text{sys}$ ranges roughly from 150\,K to 1000\,K, with a median value of 284\,K.

The incoherent gain $G(\theta,\phi)$ of the MWA for a given pointing direction can be calculated using the beam power pattern $P(\theta,\phi)$ and the offset of the target source from the zenith. Following \cite{Oronsaye2015}, we rewrite the expression of the incoherent gain as:
\begin{equation}
G(\theta,\phi) = G_\text{zenith} \times \frac{P(\theta,\phi)}{P_\text{zenith}} = \frac{\frac{\lambda^2}{2\pi}(16\sqrt{N})}{2k_{\rm B}}\times \frac{P(\theta,\phi)}{P_\text{zenith}},
\end{equation}
where $\lambda$ is the observing wavelength, $N$ is the number of MWA tiles, and $k_{\rm B}$ is the Boltzmann constant. The gain value for our detections ranges from 0.003 to 0.024\,K/Jy, with a median value of 0.013\,K/Jy.

\section{RESULTS}

In total we processed 37 hours of VCS data, and folded 1227 catalogued pulsars using our pipeline. We successfully detected 50 pulsars with $(\text{S/N})_\text{peak} \gtrsim 4$, after manually inspecting the PRESTO outputs. The detection and telescope parameters at 185\,MHz are summarised in Table \ref{tab:01}. 
The quoted errors on the flux densities ($S_{185}$) are essentially the uncertainties from our estimation of S/N and hence largely reflect the quality of our detections. However, these errors do not account for large flux density variations arising from other substantial sources of error such as the calibration technique we adopted (see section \ref{sec:flux_cal}), or those due to scintillation effects. In particular, long-term (refractive) scintillation effects can lead to flux variability of a factor of $\sim$2--3 at the MWA's frequencies \citep[e.g.][]{Gupta1993,Bhat1999a}. 
The average pulse profiles of the 50 pulsars are shown in Figure~\ref{fig:05}. For ten pulsars, with declination $\delta<-50$\degree, these are the lowest frequency detections published.

The detected pulsars have a wide range of periods (1.74\,ms--1.96\,s), DMs (2.64--180\,cm$^{-3}$\,pc), and estimated flux densities at 185\,MHz ($\sim$30--2400\,mJy).  
We detected six millisecond pulsars (MSPs): PSRs J0034$-$0534, J0437$-$4715, J1022$+$1001, J1902$-$5105, J2145$-$0750, and J2241$-$5236. Four of these are regularly monitored as part of the pulsar timing array project at Parkes \citep{Manchester2013}. 
MSPs J1902$-$5105 and J2241$-$5236 were discovered by observing unidentified \emph{Fermi} Large Area Telescope (LAT) sources using the Parkes radio telescope \citep{Keith2011,Kerr2012}, and have flux density measurements at 1.4\,GHz. The low-frequency flux density of PSR J2241$-$5236 has also been measured using continuum imaging studied with the MWA \citep[60\,mJy at 200\,MHz;][]{Murphy2017}. 
For J1902$-$5105, there are no low-frequency flux density measurements in the literature. Using MWA-VCS observations, we estimate a flux density of 362$\pm$55\,mJy at 185\,MHz. Using the pulsar's flux density measured with Parkes at 1.4\,GHz (1.2\,mJy), we calculated the spectral index $\alpha$ (assuming a single power law $S_{\nu}\propto\nu^{\alpha}$) to be approximately $-2.8$.

We also detected two non-recycled pulsars which are in binary systems, PSRs J0823$+$0159 and J1141$-$6545. The latter is in an eccentric, relativistic 4.74\,hour orbit that shows no evidence of being recycled \citep{Kaspi2000} and is an excellent laboratory for testing theories of gravity \citep{Bhat2008}. PSR J0823$+$0159, on the other hand, is in a wide-orbit (1232-day period) binary system, with a relatively low mass companion \citep{Hobbs2004}.

We were unable to obtain meaningful estimates of the mean flux density for PSRs J0534$+$2200 and J0835$-$4510 (i.e. the Crab and Vela pulsars) because their pulse profiles are severely scattered (the scattering tail extends over the entire pulsar period), and consequently the off-pulse RMS noise could not be determined reliably. PSRs J0742$-$2822, J1534$-$5334, and J1820$-$0427 also show some degree of scattering; however, their pulses are not broadened as significantly, and we were able to estimate their flux densities. More detailed analyses of scattering will be reported in a forthcoming paper (Kirsten et al. \emph{in prep.}).

\begin{figure*}[ht]
\begin{center}
\includegraphics[scale=0.75, angle=0]{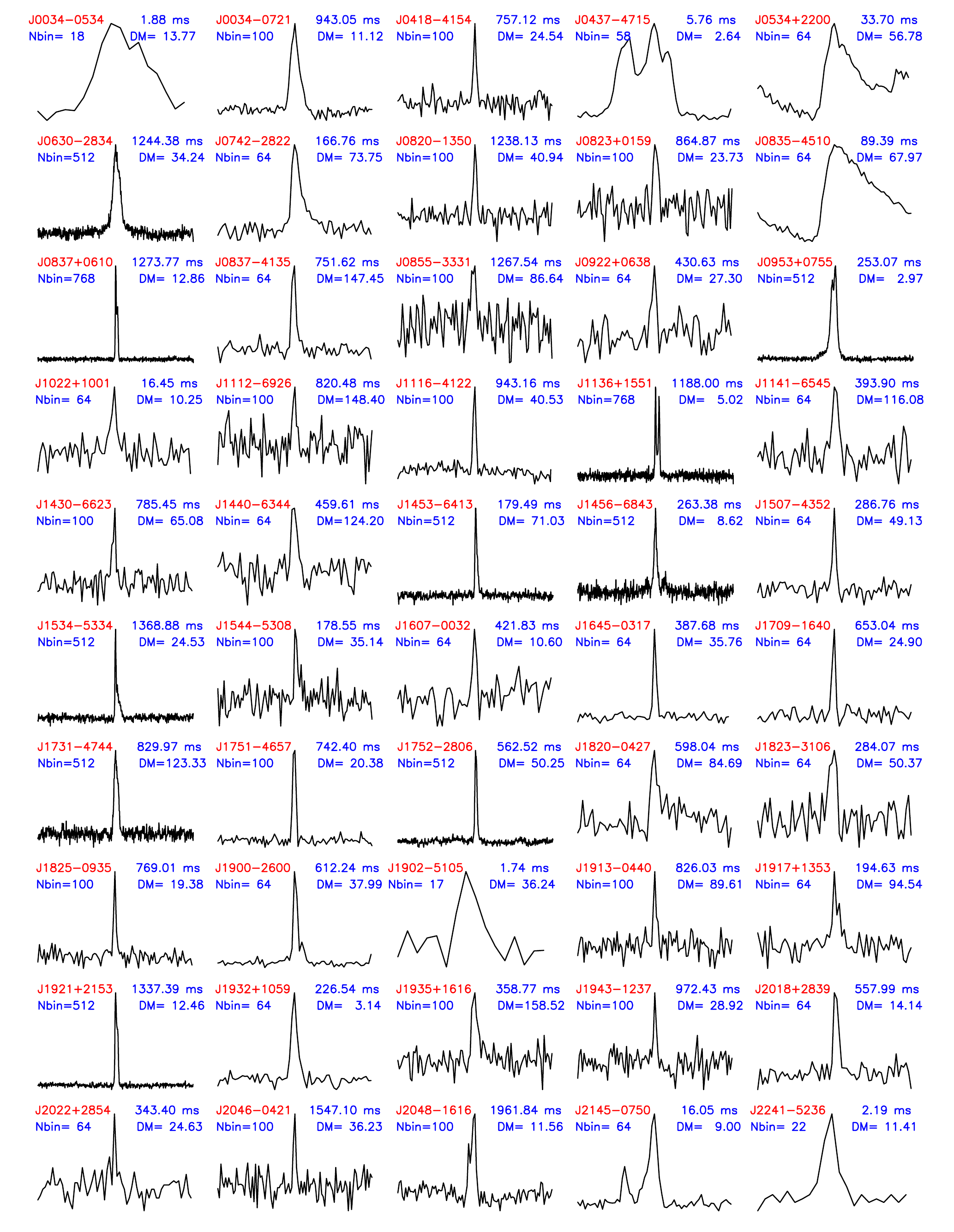}
\caption{Average pulse profiles for the 50 catalogued pulsars successfully detected using MWA-VCS incoherently-summed data at 185\,MHz, including 6 MSPs. The period, DM and number of phase bins for each pulsar are also shown. For high-S/N pulsars that we selected to present multi-frequency profiles (Figure 6), we have adopted pulse profiles of higher time resolution than as described in Section 2.4.2.}\label{fig:05}
\end{center}
\end{figure*}

\section{DISCUSSION}

Our discussion is focused on three main points. Firstly, for a subset of our pulsars, we compare our MWA profiles to those from observations made at multiple frequencies. Secondly, we estimate detection limits of this census. Finally, we review the prospects for detecting pulsars with the full sensitivities of MWA and SKA1-Low.

\subsection{Multi-frequency Pulse Profiles}

We have selected 16 pulsars with relatively high signal-to-noise (S/N$>$15) for comparison with pulse profiles across different frequencies.
The higher frequency profiles are obtained from observations using Parkes and GMRT \citep{Johnston2008,Johnston2017}, Effelsberg 100-m \citep{Hoensbroech1998, Kijak1998}, and Lovell telescopes \citep{Gould1998}. Lower frequency profiles are from recent LOFAR work \citep{LOFAR100profile2016} and the Pushchino telescope \citep{Kuzmin1999} retrieved from the EPN database\footnote{http://www.epta.eu.org/epndb/}.

Figure~\ref{fig:06} shows the evolution of the profiles with observing frequency. For pulsars located north of $\delta=-28$\degree, our MWA profiles bridge the gap between observations using LOFAR and Pushchino and those at higher frequencies. 
For nine pulsars located south of $\delta=-28$\degree, the MWA provides the lowest-frequency detections, extending the frequency range available for those pulsars by a factor of 1.3$-$7.4.

Examples of pulsars with profiles that support the RFM model include PSRs J1136$+$1551, J1752$-$2806, and J2048$-$1616. For example, the two-component profile of PSR J1136$+$1551 shows a systematic increase in pulse width and the separation of profile components towards lower frequencies. Our MWA observations are well in accordance with this trend.
PSR J2048$-$1616 has a profile with three significant components at low frequencies and the pulse width becomes narrower at higher frequencies; our MWA detection follows this trend. PSR J1752$-$2806's pulse profile has a single component that becomes narrower with increasing frequency, although this trend does not hold at the highest observing frequency, 8.4\,GHz, due to the appearance of a post-cursor component.

There are also examples of pulsars which do not follow expectations based on the RFM model.
PSR J0630$-$2834 has a single, broad component profile across all frequencies that does not show a significant change in pulse width.
Another pulsar with a two-component pulse profile, PSR J0837$+$0610, shows the components narrowing towards lower frequencies.

The multi-frequency pulse profiles for PSR J1456$-$6843 show a more complex profile evolution. The MWA profile is narrower than those at higher frequencies, possibly due to the pre- and postcursor components moving away from the main pulse and becoming less significant. 
The MWA has the ability to carry out multi-band observations (80--300\,MHz) by taking advantage of its flexible design, and this capability has been used for detailed profile and spectral studies \citep{Meyers2017}. 
Future MWA observations at frequencies below and above 185\,MHz can therefore provide further information towards investigating the profile evolution in such pulsars.

PSR J0953$+$0755 also shows a rapid profile evolution towards lower frequencies. A single component is apparent at high frequency, while a precursor evolves away from, and increases in brightness relative to, the main pulse, becoming a two-component profile below 200\,MHz. 
There are few published pulse profiles for PSR J1534$-$5334. Interestingly, scattering tails are apparent at the Parkes 1.4 and 3.1\,GHz frequencies, as well as at 185\,MHz.

MSPs J0437$-$4715 and J2145$-$0750 are detected with high S/N using the MWA.  PSR J0437$-$4715 is known to show a complex profile evolution across frequencies, where the outer components have significantly different spectra from the central components and become almost as strong as the central component at 200\,MHz \citep{Bhat2014}. PSR J2145$-$0750 has also been  detected with LOFAR \citep{LOFARmsp2016}. Further detailed study of these two MSPs will be published in an upcoming paper (Bhat et al. \emph{in prep.}). The other four MSPs detected using the incoherently-summed MWA data have limited time resolution or S/N; therefore, we do not comment further on their profiles.

\begin{figure*}[ht]
\begin{center}
\includegraphics[width=0.88\textwidth]{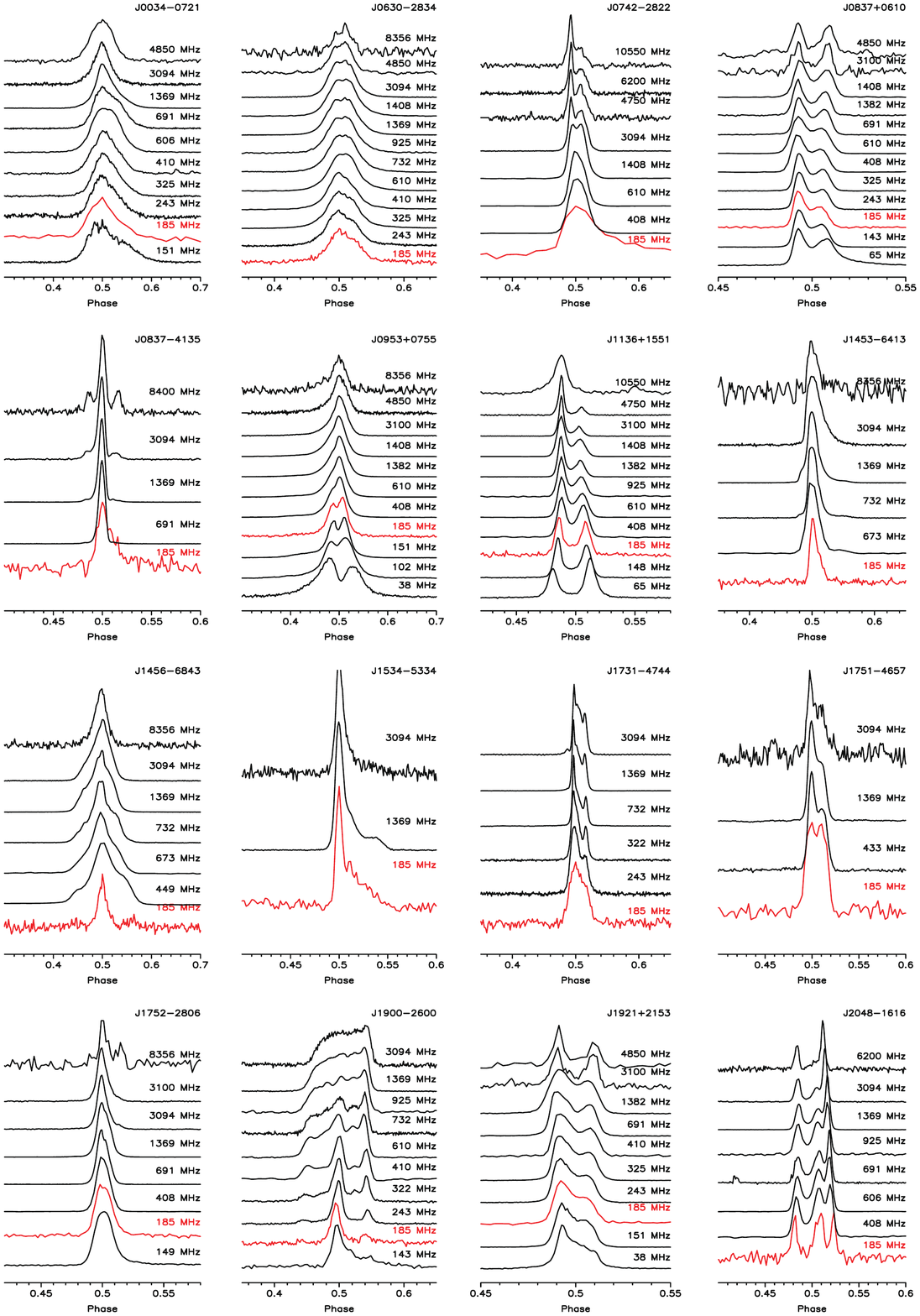}
\caption{Multi-frequency pulse profiles for 16 pulsars. MWA detections (at 185\,MHz) are shown in red. The range in pulse phase (x-axis) is chosen to suit the pulse structure and width. The profiles are normalized to the maximum value and nominally aligned based on their measured peak strengths or a suitable fiducial point. The references for the other pulse profiles in black are described in the text.
}\label{fig:06}
\end{center}
\end{figure*}

\subsection{Detection limits}\label{sec:discuss2}

To further investigate MWA's pulsar detection prospects, we summarise all our detections, as well as non detections, in Figure~\ref{fig:07}. We used the mean flux density at 400\,MHz ($S_{400}$) as an indicator of each pulsar's brightness because these measurements are the closest in frequency most widely available from the pulsar catalogue.

As described in \cite{Tingay2013}, the effective collecting area over system temperature ($A_\text{eff}/T_\text{sys}$) for a single MWA tile is around 0.1\,m$^2$/K at 185\,MHz. The noise level, $\sigma$, of MWA-VCS incoherently summed data can be expressed as:
\begin{equation}\label{eq:RMS}
\sigma=\frac{2k_\text{B}}{\sqrt{n_\text{p}t_\text{int}\Delta f N}} \frac{T_\text{sys}}{A_\text{eff}},
\end{equation}
where $t_\text{int}$ is the integration time (typically 1\,hour), and  $N$ is the number of MWA tiles incoherently summed (usually 128). Thus, the 1-$\sigma$ noise level is 5\,mJy. Recently, \cite{Murphy2017} carried out a pulsar spectral study using the MWA imaging data. Their 1-$\sigma$ RMS is 15\,mJy ${\rm beam^{-1}}$. 
The lower noise level of our data is partly due to the lack of confusion as we are using the time-variability to negate the confusion noise that limits MWA continuum imaging observations \citep{Condon1974,Franzen2016}.
Of the 50 detections in this work, PSR J0855$-$3331 has the lowest $S_{400}$, which is 7.7\,mJy. Therefore, we estimate a detection limit of $\approx 7$\,mJy for incoherently summed MWA data.

The predicted mean S/N in Figure\,\ref{fig:07}(b) is calculated using the following equation:
\begin{equation}\label{eq:S/N_pre}
  (\text{S/N})_\text{predicted}=S_{400}\left(\frac{185}{400}\right)^{-1.4}\frac{G\sqrt{n_\text{p}t_\text{int}\Delta f}}{T_\text{sys}}.
\end{equation}
To obtain the predicted S/N at 185\,MHz using the catalogue value for S$_{\rm{400}}$, we assumed a spectral index of $-1.4$ (the mean value calculated in \citealt{Bates2013}). This value was found to have a rather large standard deviation (0.96), and introduces additional uncertainty for pulsars whose spectral indices depart from this value. 
To compare the measured S/N with the $(\text{S/N})_\text{predicted}$, calculated using Eq. \eqref{eq:S/N_pre}, we scaled our measured $(\text{S/N})_\text{peak}$ to the equivalent mean S/N:

\begin{equation}\label{eq:S/N_act}
  (\text{S/N})_\text{actual}= (\text{S/N})_\text{peak}\sqrt{\frac{W}{P-W}}.
\end{equation}

Figure\,\ref{fig:07}(c) shows the comparison of $(\text{S/N})_\text{predicted}$ and $(\text{S/N})_\text{actual}$ for pulsars we detected. There is a noticeable scattering of data points from the grey dashed line that represents $\frac{(\text{S/N})_\text{{predicted}}}{(\text{S/N})_\text{{actual}}}=1$.
Possible reasons for this may include: (1) the spectral index deviation from $-1.4$, (2) the error of equivalent pulse width and the pulse width difference between 185\,MHz and 400\,MHz, (3) the flux density fluctuation caused by long-term (refractive) scintillation, and (4) uncertainty in the calibration technique we adopted. For instance, PSR J1921$+$2153 which has a $(\text{S/N})_\text{predicted}$ of 1.6 and a $(\text{S/N})_\text{actual}$ of 20 has a very steep spectral index ($\sim-3$) while the PSR J0437$-$4715 which has a $(\text{S/N})_\text{predicted}$ of 27 and a $(\text{S/N})_\text{actual}$ of 85 has a significantly large pulse width at 185\,MHz.

Of the 1227 pulsars we attempted to detect, there are 254 pulsars above the MWA incoherent sensitivity limit (i.e. $S_{400}>$7\,mJy). The large number of non-detected pulsars may be due to several reasons.
For very short period pulsars, like PSR J1939$+$2134 ($P=1.56$\,ms, $\text{DM}=71$\,cm$^{-3}$\,pc), the DM smearing within each 10\,kHz channel is 1.20\,ms (at our lowest frequency channel), which is a substantial fraction (77\%) of the pulse period. Furthermore, the pulse profile is expected to be highly scattered at our observing frequency (see \citealt{Kondratiev2016}).
Similarly, for PSR J1810$+$1744 ($P=1.66$\,ms, $\text{DM}=40$\,cm$^{-3}$\,pc), the DM smearing within one fine channel is 0.68\,ms (41\% of the pulse period). This makes detection of these pulsars difficult since we have not coherently de-dispersed our data. PSR J1902$-$5105 has the lowest $P$/DM value of all the pulsars we detected ($P$/DM$=0.048$\,ms\,cm$^{3}$\,pc$^{-1}$) which corresponds to a DM smearing of 35\% of the pulse period. However, the DM smear within a 10\,kHz fine channel can only explain a small fraction of our non-detections.

We assumed a spectral index of $-1.4$, even though it has a significant standard deviation (0.96) \citep{Bates2013}. 
Furthermore, some pulsars tend to have a flatter spectral index, or show a turn over at low frequencies, or even a broken power law spectra \citep[e.g.][]{Maron2000,Murphy2017}. 
All of these can potentially lead to non-detection. Low-frequency pulsar observations are therefore very useful in constraining the spectral behaviour of such pulsars.

Other factors influencing non-detection include the effects of scattering and scintillation. For high DM pulsars, pulse profiles become broader at lower observing frequencies, making their detection more difficult. Scintillation effects may also lead to non-detectability at low frequencies, particularly for low to moderate DM pulsars ($\text{DM}<50$\,cm$^{-3}$\,pc), whose apparent fluxes can vary by factor of $\sim$2--3, and can thus sometimes appear to be fainter than their true fluxes.

\begin{figure*}[ht]
\begin{center}
\includegraphics[scale=0.5, angle=0]{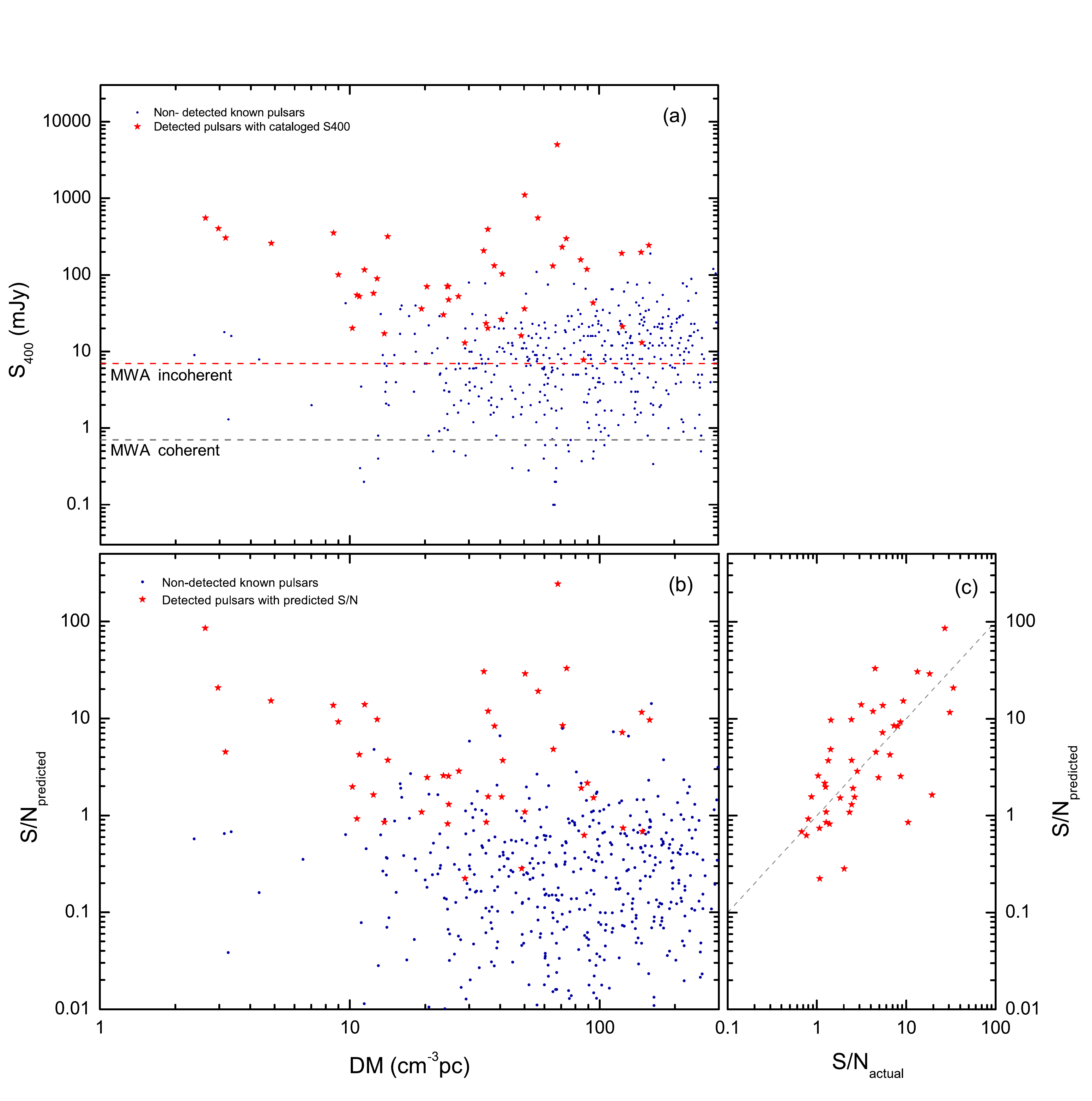}
\caption{(a). Catalogued pulsars that were folded and were or were not detected (red stars and blue points, respectively) shown in DM and $S_{400}$ parameter space. Here we use the $S_{400}$ value from the pulsar catalogue and exclude four detected pulsars that do not have a published $S_{400}$. The red dashed line shows the sensitivity limit of incoherently-summed MWA data; the grey dashed line shows the current MWA sensitivity for coherently-beamformed data.
(b). The predicted S/N against DM for pulsars that we did and did not detect (using the same symbols as in (a)). The mean S/N$_\text{predicted}$ was calculated using Eq. \eqref{eq:S/N_pre}.
(c). The predicted S/N vs. the actual S/N for pulsars we detected. Consistent with the S/N$_\text{predicted}$, the S/N$_\text{actual}$ is also the mean S/N calculated using Eq. \eqref{eq:S/N_act}. The grey dashed line represent $\frac{(\text{S/N})_\text{{predicted}}}{(\text{S/N})_\text{{actual}}}=1$
}\label{fig:07}
\end{center}
\end{figure*}

Of the 60 pulsars detected by \cite{Murphy2017} using MWA imaging data, 26 were also detected in our observations and a further 11 were not observed in VCS mode (see Table~\ref{tab:01}). Six of the 23 pulsars not detected in this work have $\text{DM}>200$\,cm$^{-3}$\,pc and consequently their profiles may be significantly scattered at 185\,MHz. The \cite{Murphy2017} detections also include the MSP J1810$+$1744 with $P=1.66$\,ms and $\text{DM}=40$\,cm$^{-3}\,$pc, for which DM smearing over the 10\,kHz channel may degrade its S/N. For the remaining 16, it is possible that long-term (refractive) scintillation effects may be a plausible reason, as such effects can give rise to flux variations of a factor of $\sim$2--3 at low frequencies \citep{Gupta1993,Bhat1999a}. 
For most pulsars in both our sample and \cite{Murphy2017} (with the exceptions of PSRs J0835$-$4510 and J0034$-$0534), the measured flux densities tend to agree within a factor of three (the flux variation that can be caused by scintillation). In the case of MSP J0034$-$0534, the measured flux density differs by a factor of six, and in the case of PSR J0835$-$4510, its long scattering tail prevented us from obtaining a meaningful estimate of its flux.

\subsection{Future detection prospects}\label{sec:discuss3}

We use the results from our initial census to examine the prospects of detecting pulsars using the full sensitivity of MWA (coherently beamformed data) as well as using the SKA1-Low, i.e. Phase I of SKA-Low.
As discussed in section\,\ref{sec:discuss2}, the estimated sensitivity limit for the incoherent beam is $\approx 7$\,mJy at 400\,MHz.
The sensitivity is expected to improve by a factor of $\sqrt{N_\text{tile}}\approx$11 for coherently beamformed MWA-VCS data. However, in reality, this will depend on a number of factors such as calibration accuracy, the number of tiles included, the source and background temperature, and potentially even correlated noise between adjacent tiles.
For example, \citet{Bhat2016} report a factor of 10 improvement for their observations of MSP J0437$-$4715. This is nearly 90\% of the theoretical expectation.
We therefore assume the theoretical sensitivity limit for the incoherently summed data to be $7$\,mJy, and that for the coherently beamformed data to be 10 times more sensitive, $0.7$\,mJy, as shown in Figures\,\ref{fig:07} and\,\ref{fig:08}.

SKA1-Low is planned to consist of around \,130,000 log-periodic dual-polarised antenna elements, designed for sensitivity from 50 to 350 MHz.
From Fig.\,19 of the SKA Baseline Design document v2, $A_\text{eff}/T_\text{sys}$ for SKA1-Low is around \,600\,m$^2$/K at 200\,MHz (towards zenith) and the corresponding $T_\text{sys}$ is 170\,K. Thus, the gain at 200\,MHz can be calculated as $G=A_\text{eff}/2k_\text{B}\approx37$\,m$^2$/K. We compare the
sensitivity limit for SKA1-Low with that for the incoherently summed MWA (see Figure\,\ref{fig:08}). $A_\text{eff}/T_\text{sys}$ for SKA1-Low is approximately 600 times better than the MWA incoherent array (which has $A_\text{eff}/T_\text{sys} \sim 1$\,m$^2$/K; \citealt{Tingay2013}). The sensitivity will further improve by a factor of $\sqrt{3}$ assuming a bandwidth of 100\,MHz for SKA1-Low\footnote{Pulsar flux densities, dispersion smearing, scattering broadening and $T_\text{sky}$ all vary across the large observing frequency range and we do not explicitly account for these effects.}. 
By factoring in shorter integration times (600\,s, i.e. 1/6 of that used for the MWA sensitivity calculation), we estimate the sensitivity limit of SKA1-Low to be around 400 times lower than that of the MWA incoherently summed data.

To gain further insight into the pulsar detection prospects of SKA1-Low, we simulated pulsar populations using the PsrPopPy \citep{Bates2014} code. 
New functionality has been implemented in PsrPopPy, which calculates the telescope gain as a cosine-square function of the zenith distance, providing more realistic pulsar yields for an aperture array.
Using our MWA incoherent census as an input survey with zenith gain G=0.025 (maximum gain value in Table\,\ref{tab:01}) and considering only pulsars with $(\text{S/N})_\text{peak}>$5, the corresponding detection count is 57.
This includes 9 more detections from recent observations (whose profiles are not reported here as the data are still within the proprietary period and will be published separately)\footnote{Of the 50 detections we report in this paper, two have  $(\text{S/N})_\text{peak}$ $\sim$4 and therefore excluded from this analysis.}.
We simulated 120 realisations of pulsar populations to span a wide range of population characteristics (e.g. luminosity distribution, spatial distribution, etc.).
For each realisation, we performed 120 survey simulations at a frequency of 200\,MHz assuming a bandwidth of 100\,MHz and an integration time of 600\,s. Thus, we ran 14,400 simulated SKA1-Low surveys in total. The mean value of pulsars detected in this simulated SKA1-Low survey is 9431$\pm$1279, which is over 3.5-times the number of known pulsars to date. We note that the spectral turnover at low-frequencies will also affect the pulsar detection prospects for SKA1-Low, though it is not well characterised for most pulsars.

The predicted SKA1-Low yields from one randomly chosen PsrPopPy simulation (10,409 pulsars detected in this simulation) are plotted in Figure\,\ref{fig:08}. To cross-check these results, we ran a simulation of an MWA survey on this specific population model, which predicted 56 detections (also shown in Figure\,\ref{fig:08}). We note that the PsrPopPy detections have a very similar $S_{400}$ distribution while the DM distribution extends to lower DM values when compared with the pulsar catalogue. Our results are comparable to those from \cite{Keane2015}, who also found the optimum frequency for pulsar searching is 250\,MHz for SKA1-Low. For comparison, they predict SKA1-Low will be able to detect $\sim$ 7000 normal pulsars and $\sim$ 900 MSPs.

\begin{figure*}[ht]
\begin{center}
\includegraphics[scale=0.6, angle=0]{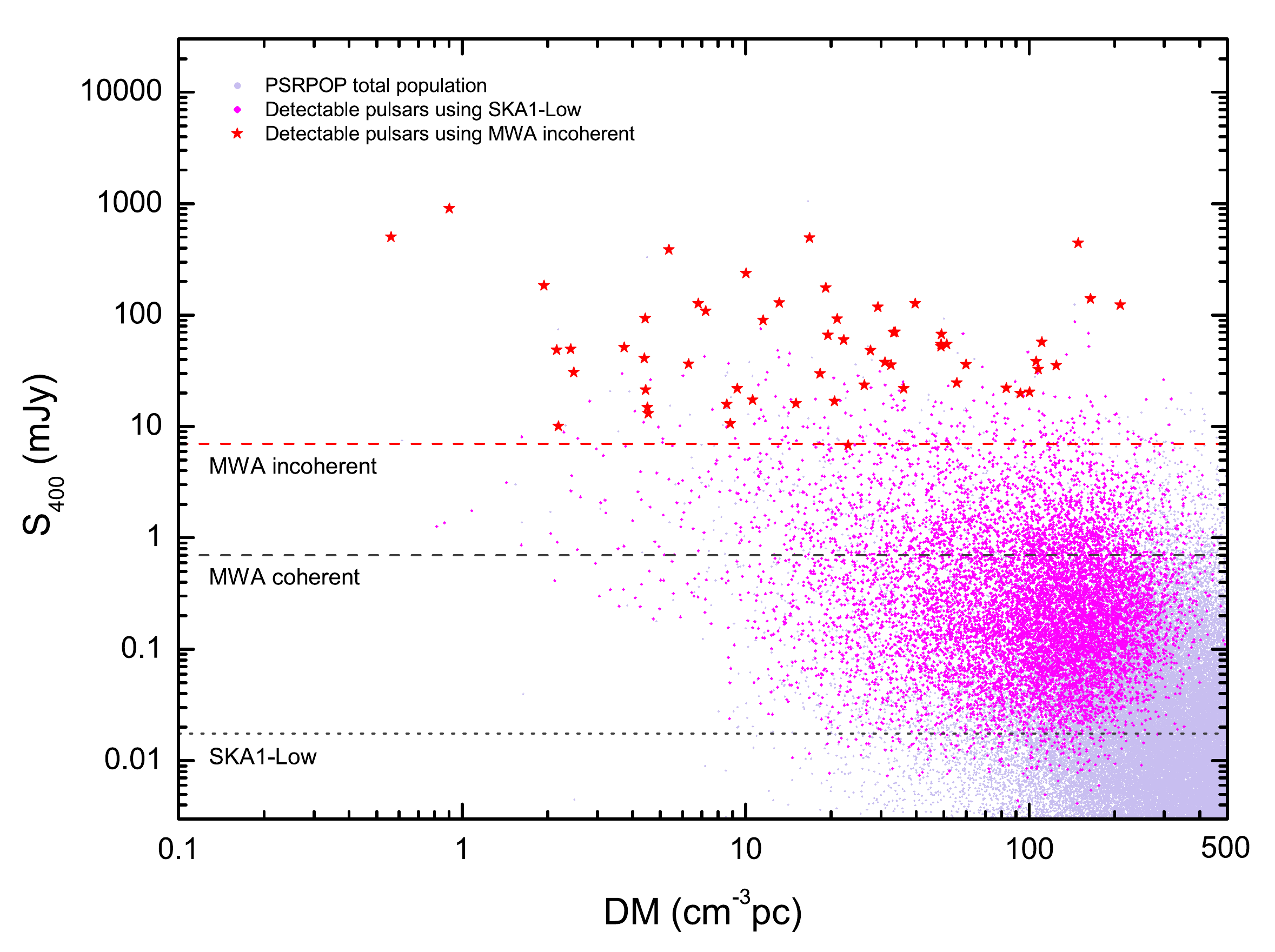}
\caption{An example of a simulated pulsar population and predicted pulsar detections using the MWA (incoherent sum) and SKA1-Low.
The light grey points indicate the total pulsar population generated by PsrPopPy using MWA incoherent-sum parameters and a detection number of 57.
The red stars indicate the predicted detections from an MWA incoherent survey.
The magenta dots indicate the predicted detections from an SKA1-Low survey.
The red and grey dashed lines indicate the sensitivity limits for the incoherently and coherently summed MWA data, respectively (identical to Figure\,\ref{fig:07}a).
The grey dotted line shows the estimated sensitivity limit of SKA1-Low.
}\label{fig:08}

\end{center}
\end{figure*}

\section{SUMMARY}
We have carried out a census of known southern pulsars using 37 hours of MWA-VCS archival data from observations at 185\,MHz. This is a relatively shallow census as the data from all 128 MWA tiles were combined incoherently, yielding $\sim$10\% of the sensitivity achievable with the MWA. However, this preserves the large field-of-view ($\sim$450\,deg$^2$ at 185\,MHz) and thus facilitates an expedited initial census. We successfully detected 50 pulsars, 6 of which are millisecond pulsars. For ten pulsars, we present the lowest frequency detections available in the literature. For a subset of the pulsars we also present their multi-frequency pulse profiles by combining our data with those from other telescopes, demonstrating a number of profile evolution behaviours with frequency. We use our results to forecast a pulsar survey yield of $\sim$9400 using SKA1-Low.

\begin{acknowledgements}

This scientific work makes use of the Murchison Radio-astronomy Observatory, operated by CSIRO. We acknowledge the Wajarri Yamatji people as the traditional owners of the Observatory site. Support for the operation of the MWA is provided by the Australian Government (NCRIS), under a contract to Curtin University administered by Astronomy Australia Limited. We acknowledge the Pawsey Supercomputing Centre which is supported by the Western Australian and Australian Governments. We thank Tara Murphy for her thorough reading and valuable suggestions of this paper. This research was conducted by the Australian Research Council Centre of Excellence for All-sky Astrophysics (CAASTRO), through project number CE110001020. M. Xue is funded by China Scholarship Council through the SKA PhD Scholarship project. N. D. R. B. acknowledges the support from a Curtin Research Fellowship (CRF12228). B.W.M and S.J.M. acknowledge the support from Australian Government Research Training Program Scholarship. D.L.K. is additionally supported by NSF grant AST-1412421. 

\end{acknowledgements}

\bibliographystyle{pasa-mnras}
\bibliography{pulsar}

\begin{thebibliography}{}
\makeatletter
\relax
\def\mn@urlcharsother{\let\do\@makeother \do\$\do\&\do\#\do\^\do\_\do\%\do\~}
\definecolor{darkblue}{rgb}{0,0,0.597656}
\def\mndoi{\begingroup\mn@urlcharsother \@ifnextchar [ {\mndoi@} {\mndoi@[]}}
\def\mndoi@[#1]#2{\def\@tempa{#1}\ifx\@tempa\@empty \href
  {http://dx.doi.org/#2} {\textcolor{darkblue}{doi:#2}}\else \href
  {http://dx.doi.org/#2} {\textcolor{darkblue}{#1}}\fi \endgroup}
\def\mn@eprint#1#2{\mn@eprint@#1:#2::\@nil}
\def\mn@eprint@arXiv#1{\href {http://arxiv.org/abs/#1} {{\tt arXiv:#1}}}
\def\mn@eprint@dblp#1{\href {http://dblp.uni-trier.de/rec/bibtex/#1.xml}
  {dblp:#1}}
\def\mn@eprint@#1:#2:#3:#4\@nil{\def\@tempa {#1}\def\@tempb {#2}\def\@tempc
  {#3}\ifx \@tempc \@empty \let \@tempc \@tempb \let \@tempb \@tempa \fi \ifx
  \@tempb \@empty \def\@tempb {arXiv}\fi \@ifundefined
  {mn@eprint@\@tempb}{\@tempb:\@tempc}{\expandafter \expandafter \csname
  mn@eprint@\@tempb\endcsname \expandafter{\@tempc}}}

\bibitem[\protect\citeauthoryear{{Bates}, {Lorimer}  \& {Verbiest}}{{Bates}
  et~al.}{2013}]{Bates2013}
{Bates} S.~D.,  {Lorimer} D.~R.,   {Verbiest} J.~P.~W.,  2013, \mndoi [\mnras]
  {10.1093/mnras/stt257}, \href
  {http://adsabs.harvard.edu/abs/2013MNRAS.431.1352B} {431, 1352}

\bibitem[\protect\citeauthoryear{{Bates}, {Lorimer}, {Rane}  \&
  {Swiggum}}{{Bates} et~al.}{2014}]{Bates2014}
{Bates} S.~D.,  {Lorimer} D.~R.,  {Rane} A.,   {Swiggum} J.,  2014, \mndoi
  [\mnras] {10.1093/mnras/stu157}, \href
  {http://adsabs.harvard.edu/abs/2014MNRAS.439.2893B} {439, 2893}

\bibitem[\protect\citeauthoryear{{Bhat}, {Rao}  \& {Gupta}}{{Bhat}
  et~al.}{1999}]{Bhat1999a}
{Bhat} N.~D.~R.,  {Rao} A.~P.,   {Gupta} Y.,  1999, \mndoi [\apjs]
  {10.1086/313198}, \href {http://adsabs.harvard.edu/abs/1999ApJS..121..483B}
  {121, 483}

\bibitem[\protect\citeauthoryear{{Bhat}, {Cordes}, {Camilo}, {Nice}  \&
  {Lorimer}}{{Bhat} et~al.}{2004}]{Bhat2004}
{Bhat} N.~D.~R.,  {Cordes} J.~M.,  {Camilo} F.,  {Nice} D.~J.,   {Lorimer}
  D.~R.,  2004, \mndoi [\apj] {10.1086/382680}, \href
  {http://adsabs.harvard.edu/abs/2004ApJ...605..759B} {605, 759}

\bibitem[\protect\citeauthoryear{{Bhat}, {Bailes}  \& {Verbiest}}{{Bhat}
  et~al.}{2008}]{Bhat2008}
{Bhat} N.~D.~R.,  {Bailes} M.,   {Verbiest} J.~P.~W.,  2008, \mndoi [\prd]
  {10.1103/PhysRevD.77.124017}, \href
  {http://adsabs.harvard.edu/abs/2008PhRvD..77l4017B} {77, 124017}

\bibitem[\protect\citeauthoryear{{Bhat} et~al.,}{{Bhat}
  et~al.}{2014}]{Bhat2014}
{Bhat} N.~D.~R.,  et~al., 2014, \mndoi [\apjl] {10.1088/2041-8205/791/2/L32},
  \href {http://adsabs.harvard.edu/abs/2014ApJ...791L..32B} {791, L32}

\bibitem[\protect\citeauthoryear{{Bhat}, {Ord}, {Tremblay}, {McSweeney}  \&
  {Tingay}}{{Bhat} et~al.}{2016}]{Bhat2016}
{Bhat} N.~D.~R.,  {Ord} S.~M.,  {Tremblay} S.~E.,  {McSweeney} S.~J.,
  {Tingay} S.~J.,  2016, \mndoi [\apj] {10.3847/0004-637X/818/1/86}, \href
  {http://adsabs.harvard.edu/abs/2016ApJ...818...86B} {818, 86}

\bibitem[\protect\citeauthoryear{{Bilous} et~al.,}{{Bilous}
  et~al.}{2016}]{Bilous:2016}
{Bilous} A.~V.,  et~al., 2016, \mndoi [\aap] {10.1051/0004-6361/201527702},
  \href {http://adsabs.harvard.edu/abs/2016A%26A...591A.134B} {591, A134}

\bibitem[\protect\citeauthoryear{{Condon}}{{Condon}}{1974}]{Condon1974}
{Condon} J.~J.,  1974, \mndoi [\apj] {10.1086/152714}, \href
  {http://adsabs.harvard.edu/abs/1974ApJ...188..279C} {188, 279}

\bibitem[\protect\citeauthoryear{{Cordes}}{{Cordes}}{1978}]{Cordes1978}
{Cordes} J.~M.,  1978, \mndoi [\apj] {10.1086/156218}, \href
  {http://ads.bao.ac.cn/abs/1978ApJ...222.1006C} {222, 1006}

\bibitem[\protect\citeauthoryear{{Cordes} \& {Lazio}}{{Cordes} \&
  {Lazio}}{2002}]{NE2001}
{Cordes} J.~M.,  {Lazio} T.~J.~W.,  2002, ArXiv Astrophysics e-prints, \href
  {http://adsabs.harvard.edu/abs/2002astro.ph..7156C} {}

\bibitem[\protect\citeauthoryear{{Dolch} et~al.,}{{Dolch}
  et~al.}{2014}]{Dolch:2014}
{Dolch} T.,  et~al., 2014, \mndoi [\apj] {10.1088/0004-637X/794/1/21}, \href
  {http://adsabs.harvard.edu/abs/2014ApJ...794...21D} {794, 21}

\bibitem[\protect\citeauthoryear{{Franzen} et~al.,}{{Franzen}
  et~al.}{2016}]{Franzen2016}
{Franzen} T.~M.~O.,  et~al., 2016, \mndoi [\mnras] {10.1093/mnras/stw823},
  \href {http://adsabs.harvard.edu/abs/2016MNRAS.459.3314F} {459, 3314}

\bibitem[\protect\citeauthoryear{{Gould} \& {Lyne}}{{Gould} \&
  {Lyne}}{1998}]{Gould1998}
{Gould} D.~M.,  {Lyne} A.~G.,  1998, \mndoi [\mnras]
  {10.1046/j.1365-8711.1998.02018.x}, \href
  {http://adsabs.harvard.edu/abs/1998MNRAS.301..235G} {301, 235}

\bibitem[\protect\citeauthoryear{{Gupta}, {Rickett}  \& {Coles}}{{Gupta}
  et~al.}{1993}]{Gupta1993}
{Gupta} Y.,  {Rickett} B.~J.,   {Coles} W.~A.,  1993, \mndoi [\apj]
  {10.1086/172193}, \href {http://adsabs.harvard.edu/abs/1993ApJ...403..183G}
  {403, 183}

\bibitem[\protect\citeauthoryear{{Haslam}, {Klein}, {Salter}, {Stoffel},
  {Wilson}, {Cleary}, {Cooke}  \& {Thomasson}}{{Haslam}
  et~al.}{1981}]{Haslam1981}
{Haslam} C.~G.~T.,  {Klein} U.,  {Salter} C.~J.,  {Stoffel} H.,  {Wilson}
  W.~E.,  {Cleary} M.~N.,  {Cooke} D.~J.,   {Thomasson} P.,  1981, \aap, \href
  {http://adsabs.harvard.edu/abs/1981A%26A...100..209H} {100, 209}

\bibitem[\protect\citeauthoryear{{Haslam}, {Salter}, {Stoffel}  \&
  {Wilson}}{{Haslam} et~al.}{1982}]{Haslam1982}
{Haslam} C.~G.~T.,  {Salter} C.~J.,  {Stoffel} H.,   {Wilson} W.~E.,  1982,
  \aaps, \href {http://adsabs.harvard.edu/abs/1982A%26AS...47....1H} {47, 1}

\bibitem[\protect\citeauthoryear{{Hassall} et~al.,}{{Hassall}
  et~al.}{2012}]{Hassall:2012}
{Hassall} T.~E.,  et~al., 2012, \mndoi [\aap] {10.1051/0004-6361/201218970},
  \href {http://adsabs.harvard.edu/abs/2012A%26A...543A..66H} {543, A66}

\bibitem[\protect\citeauthoryear{{Hermsen} et~al.,}{{Hermsen}
  et~al.}{2013}]{Hermsen2013}
{Hermsen} W.,  et~al., 2013, \mndoi [Science] {10.1126/science.1230960}, \href
  {http://adsabs.harvard.edu/abs/2013Sci...339..436H} {339, 436}

\bibitem[\protect\citeauthoryear{{Hewish}, {Bell}, {Pilkington}, {Scott}  \&
  {Collins}}{{Hewish} et~al.}{1968}]{Hewish1968}
{Hewish} A.,  {Bell} S.~J.,  {Pilkington} J.~D.~H.,  {Scott} P.~F.,   {Collins}
  R.~A.,  1968, \mndoi [\nat] {10.1038/217709a0}, \href
  {http://adsabs.harvard.edu/abs/1968Natur.217..709H} {217, 709}

\bibitem[\protect\citeauthoryear{{Hobbs}, {Lyne}, {Kramer}, {Martin}  \&
  {Jordan}}{{Hobbs} et~al.}{2004}]{Hobbs2004}
{Hobbs} G.,  {Lyne} A.~G.,  {Kramer} M.,  {Martin} C.~E.,   {Jordan} C.,  2004,
  \mndoi [\mnras] {10.1111/j.1365-2966.2004.08157.x}, \href
  {http://adsabs.harvard.edu/abs/2004MNRAS.353.1311H} {353, 1311}

\bibitem[\protect\citeauthoryear{{Hotan}, {van Straten}  \&
  {Manchester}}{{Hotan} et~al.}{2004}]{Hotan2004}
{Hotan} A.~W.,  {van Straten} W.,   {Manchester} R.~N.,  2004, \mndoi [\pasa]
  {10.1071/AS04022}, \href {http://adsabs.harvard.edu/abs/2004PASA...21..302H}
  {21, 302}

\bibitem[\protect\citeauthoryear{{Howard}, {Stovall}, {Dowell}, {Taylor}  \&
  {White}}{{Howard} et~al.}{2016}]{Howard:2016}
{Howard} T.~A.,  {Stovall} K.,  {Dowell} J.,  {Taylor} G.~B.,   {White} S.~M.,
  2016, \mndoi [\apj] {10.3847/0004-637X/831/2/208}, \href
  {http://adsabs.harvard.edu/abs/2016ApJ...831..208H} {831, 208}

\bibitem[\protect\citeauthoryear{{Johnston} \& {Kerr}}{{Johnston} \&
  {Kerr}}{2018}]{Johnston2017}
{Johnston} S.,  {Kerr} M.,  2018, \mndoi [\mnras] {10.1093/mnras/stx3095},
  \href {http://adsabs.harvard.edu/abs/2018MNRAS.474.4629J} {474, 4629}

\bibitem[\protect\citeauthoryear{{Johnston}, {Karastergiou}, {Mitra}  \&
  {Gupta}}{{Johnston} et~al.}{2008}]{Johnston2008}
{Johnston} S.,  {Karastergiou} A.,  {Mitra} D.,   {Gupta} Y.,  2008, \mndoi
  [\mnras] {10.1111/j.1365-2966.2008.13379.x}, \href
  {http://adsabs.harvard.edu/abs/2008MNRAS.388..261J} {388, 261}

\bibitem[\protect\citeauthoryear{{Kaspi} et~al.,}{{Kaspi}
  et~al.}{2000}]{Kaspi2000}
{Kaspi} V.~M.,  et~al., 2000, \mndoi [\apj] {10.1086/317103}, \href
  {http://adsabs.harvard.edu/abs/2000ApJ...543..321K} {543, 321}

\bibitem[\protect\citeauthoryear{{Keane} et~al.,}{{Keane}
  et~al.}{2015}]{Keane2015}
{Keane} E.,  et~al., 2015, Advancing Astrophysics with the Square Kilometre
  Array (AASKA14), \href {http://adsabs.harvard.edu/abs/2015aska.confE..40K}
  {p.~40}

\bibitem[\protect\citeauthoryear{{Keith} et~al.,}{{Keith}
  et~al.}{2011}]{Keith2011}
{Keith} M.~J.,  et~al., 2011, \mndoi [\mnras]
  {10.1111/j.1365-2966.2011.18464.x}, \href
  {http://adsabs.harvard.edu/abs/2011MNRAS.414.1292K} {414, 1292}

\bibitem[\protect\citeauthoryear{{Kerr} et~al.,}{{Kerr}
  et~al.}{2012}]{Kerr2012}
{Kerr} M.,  et~al., 2012, \mndoi [\apjl] {10.1088/2041-8205/748/1/L2}, \href
  {http://adsabs.harvard.edu/abs/2012ApJ...748L...2K} {748, L2}

\bibitem[\protect\citeauthoryear{{Kijak} \& {Gil}}{{Kijak} \&
  {Gil}}{1998}]{Kijak1998a}
{Kijak} J.,  {Gil} J.,  1998, \mndoi [\mnras]
  {10.1046/j.1365-8711.1998.01832.x}, \href
  {http://adsabs.harvard.edu/abs/1998MNRAS.299..855K} {299, 855}

\bibitem[\protect\citeauthoryear{{Kijak} \& {Gil}}{{Kijak} \&
  {Gil}}{2003}]{Kijak2003}
{Kijak} J.,  {Gil} J.,  2003, \mndoi [\aap] {10.1051/0004-6361:20021583}, \href
  {http://adsabs.harvard.edu/abs/2003A%26A...397..969K} {397, 969}

\bibitem[\protect\citeauthoryear{{Kijak}, {Kramer}, {Wielebinski}  \&
  {Jessner}}{{Kijak} et~al.}{1998}]{Kijak1998}
{Kijak} J.,  {Kramer} M.,  {Wielebinski} R.,   {Jessner} A.,  1998, \mndoi
  [\aaps] {10.1051/aas:1998340}, \href
  {http://adsabs.harvard.edu/abs/1998A%26AS..127..153K} {127, 153}

\bibitem[\protect\citeauthoryear{{Kondratiev} et~al.,}{{Kondratiev}
  et~al.}{2016a}]{LOFARmsp2016}
{Kondratiev} V.~I.,  et~al., 2016a, \mndoi [\aap]
  {10.1051/0004-6361/201527178}, \href
  {http://adsabs.harvard.edu/abs/2016A%26A...585A.128K} {585, A128}

\bibitem[\protect\citeauthoryear{{Kondratiev} et~al.,}{{Kondratiev}
  et~al.}{2016b}]{Kondratiev2016}
{Kondratiev} V.~I.,  et~al., 2016b, \mndoi [\aap]
  {10.1051/0004-6361/201527178}, \href
  {http://adsabs.harvard.edu/abs/2016A%26A...585A.128K} {585, A128}

\bibitem[\protect\citeauthoryear{{Kuz'min} \& {Losovskii}}{{Kuz'min} \&
  {Losovskii}}{1999}]{Kuzmin1999}
{Kuz'min} A.~D.,  {Losovskii} B.~Y.,  1999, Astronomy Reports, \href
  {http://adsabs.harvard.edu/abs/1999ARep...43..288K} {43, 288}

\bibitem[\protect\citeauthoryear{{Lawson}, {Mayer}, {Osborne}  \&
  {Parkinson}}{{Lawson} et~al.}{1987}]{Lawson:1987}
{Lawson} K.~D.,  {Mayer} C.~J.,  {Osborne} J.~L.,   {Parkinson} M.~L.,  1987,
  \mndoi [\mnras] {10.1093/mnras/225.2.307}, \href
  {http://adsabs.harvard.edu/abs/1987MNRAS.225..307L} {225, 307}

\bibitem[\protect\citeauthoryear{Lorimer \& Kramer}{Lorimer \&
  Kramer}{2005}]{Lorimer2005handbook}
Lorimer D.,  Kramer M.,  2005, Handbook of Pulsar Astronomy.
Cambridge Observing Handbooks for Research Astronomers, Cambridge University
  Press, \url {https://books.google.com.au/books?id=OZ8tdN6qJcsC}

\bibitem[\protect\citeauthoryear{{Manchester}, {Hobbs}, {Teoh}  \&
  {Hobbs}}{{Manchester} et~al.}{2005}]{PSRCAT2005}
{Manchester} R.~N.,  {Hobbs} G.~B.,  {Teoh} A.,   {Hobbs} M.,  2005, \mndoi
  [\aj] {10.1086/428488}, \href
  {http://adsabs.harvard.edu/abs/2005AJ....129.1993M} {129, 1993}

\bibitem[\protect\citeauthoryear{{Manchester} et~al.,}{{Manchester}
  et~al.}{2013}]{Manchester2013}
{Manchester} R.~N.,  et~al., 2013, \mndoi [\pasa] {10.1017/pasa.2012.017},
  \href {http://adsabs.harvard.edu/abs/2013PASA...30...17M} {30, e017}

\bibitem[\protect\citeauthoryear{{Maron}, {Kijak}, {Kramer}  \&
  {Wielebinski}}{{Maron} et~al.}{2000}]{Maron2000}
{Maron} O.,  {Kijak} J.,  {Kramer} M.,   {Wielebinski} R.,  2000, \mndoi
  [\aaps] {10.1051/aas:2000298}, \href
  {http://adsabs.harvard.edu/abs/2000A%26AS..147..195M} {147, 195}

\bibitem[\protect\citeauthoryear{{Mereghetti} et~al.,}{{Mereghetti}
  et~al.}{2016}]{Mereghetti2016}
{Mereghetti} S.,  et~al., 2016, \mndoi [\apj] {10.3847/0004-637X/831/1/21},
  \href {http://adsabs.harvard.edu/abs/2016ApJ...831...21M} {831, 21}

\bibitem[\protect\citeauthoryear{{Meyers} et~al.,}{{Meyers}
  et~al.}{2017}]{Meyers2017}
{Meyers} B.~W.,  et~al., 2017, \mndoi [\apj] {10.3847/1538-4357/aa8bba}, \href
  {http://adsabs.harvard.edu/abs/2017ApJ...851...20M} {851, 20}

\bibitem[\protect\citeauthoryear{{Mitra} \& {Rankin}}{{Mitra} \&
  {Rankin}}{2002}]{Mitra&Rankin2002}
{Mitra} D.,  {Rankin} J.~M.,  2002, \mndoi [\apj] {10.1086/342136}, \href
  {http://adsabs.harvard.edu/abs/2002ApJ...577..322M} {577, 322}

\bibitem[\protect\citeauthoryear{{Murphy} et~al.,}{{Murphy}
  et~al.}{2017}]{Murphy2017}
{Murphy} T.,  et~al., 2017, \mndoi [\pasa] {10.1017/pasa.2017.13}, \href
  {http://adsabs.harvard.edu/abs/2017PASA...34...20M} {34, e020}

\bibitem[\protect\citeauthoryear{{Noutsos} et~al.,}{{Noutsos}
  et~al.}{2015}]{Noutsos:2015}
{Noutsos} A.,  et~al., 2015, \mndoi [\aap] {10.1051/0004-6361/201425186}, \href
  {http://adsabs.harvard.edu/abs/2015A%26A...576A..62N} {576, A62}

\bibitem[\protect\citeauthoryear{{Offringa} et~al.,}{{Offringa}
  et~al.}{2015}]{Offringa2015}
{Offringa} A.~R.,  et~al., 2015, \mndoi [\pasa] {10.1017/pasa.2015.7}, \href
  {http://adsabs.harvard.edu/abs/2015PASA...32....8O} {32, e008}

\bibitem[\protect\citeauthoryear{{Oronsaye} et~al.,}{{Oronsaye}
  et~al.}{2015}]{Oronsaye2015}
{Oronsaye} S.~I.,  et~al., 2015, \mndoi [\apj] {10.1088/0004-637X/809/1/51},
  \href {http://adsabs.harvard.edu/abs/2015ApJ...809...51O} {809, 51}

\bibitem[\protect\citeauthoryear{{Pilia} et~al.,}{{Pilia}
  et~al.}{2016}]{LOFAR100profile2016}
{Pilia} M.,  et~al., 2016, \mndoi [\aap] {10.1051/0004-6361/201425196}, \href
  {http://adsabs.harvard.edu/abs/2016A%26A...586A..92P} {586, A92}

\bibitem[\protect\citeauthoryear{{Prabu} et~al.,}{{Prabu}
  et~al.}{2015}]{Prabu2015}
{Prabu} T.,  et~al., 2015, \mndoi [Experimental Astronomy]
  {10.1007/s10686-015-9444-3}, \href
  {http://adsabs.harvard.edu/abs/2015ExA....39...73P} {39, 73}

\bibitem[\protect\citeauthoryear{{Ransom}}{{Ransom}}{2001}]{Ransom2001}
{Ransom} S.~M.,  2001, PhD thesis, Harvard University

\bibitem[\protect\citeauthoryear{{Roy}, {Gupta}, {Pen}, {Peterson}, {Kudale}
  \& {Kodilkar}}{{Roy} et~al.}{2010}]{Roy2010ExA}
{Roy} J.,  {Gupta} Y.,  {Pen} U.-L.,  {Peterson} J.~B.,  {Kudale} S.,
  {Kodilkar} J.,  2010, \mndoi [Experimental Astronomy]
  {10.1007/s10686-010-9187-0}, \href
  {http://adsabs.harvard.edu/abs/2010ExA....28...25R} {28, 25}

\bibitem[\protect\citeauthoryear{{Sokolowski} et~al.,}{{Sokolowski}
  et~al.}{2017}]{Sokolowski2017}
{Sokolowski} M.,  et~al., 2017, \mndoi [\pasa] {10.1017/pasa.2017.54}, \href
  {http://adsabs.harvard.edu/abs/2017PASA...34...62S} {34, e062}

\bibitem[\protect\citeauthoryear{{Stappers} et~al.,}{{Stappers}
  et~al.}{2011}]{Stappers2011}
{Stappers} B.~W.,  et~al., 2011, \mndoi [\aap] {10.1051/0004-6361/201116681},
  \href {http://adsabs.harvard.edu/abs/2011A%26A...530A..80S} {530, A80}

\bibitem[\protect\citeauthoryear{{Stovall} et~al.,}{{Stovall}
  et~al.}{2015}]{Stovall2015}
{Stovall} K.,  et~al., 2015, \mndoi [\apj] {10.1088/0004-637X/808/2/156}, \href
  {http://adsabs.harvard.edu/abs/2015ApJ...808..156S} {808, 156}

\bibitem[\protect\citeauthoryear{{Sutinjo}, {O'Sullivan}, {Lenc}, {Wayth},
  {Padhi}, {Hall}  \& {Tingay}}{{Sutinjo} et~al.}{2015}]{MWAbeam2015}
{Sutinjo} A.,  {O'Sullivan} J.,  {Lenc} E.,  {Wayth} R.~B.,  {Padhi} S.,
  {Hall} P.,   {Tingay} S.~J.,  2015, \mndoi [Radio Science]
  {10.1002/2014RS005517}, \href
  {http://adsabs.harvard.edu/abs/2015RaSc...50...52S} {50, 52}

\bibitem[\protect\citeauthoryear{{Swarup}, {Ananthakrishnan}, {Kapahi}, {Rao},
  {Subrahmanya}  \& {Kulkarni}}{{Swarup} et~al.}{1991}]{GMRT1991}
{Swarup} G.,  {Ananthakrishnan} S.,  {Kapahi} V.~K.,  {Rao} A.~P.,
  {Subrahmanya} C.~R.,   {Kulkarni} V.~K.,  1991, Current Science, Vol.~60,
  NO.2/JAN25, P.~95, 1991, \href
  {http://adsabs.harvard.edu/abs/1991CuSc...60...95S} {60, 95}

\bibitem[\protect\citeauthoryear{{Taylor} \& {Manchester}}{{Taylor} \&
  {Manchester}}{1977}]{Taylor1977}
{Taylor} J.~H.,  {Manchester} R.~N.,  1977, \mndoi [\araa]
  {10.1146/annurev.aa.15.090177.000315}, \href
  {http://adsabs.harvard.edu/abs/1977ARA%26A..15...19T} {15, 19}

\bibitem[\protect\citeauthoryear{{Taylor} \& {Stinebring}}{{Taylor} \&
  {Stinebring}}{1986}]{Taylor1986}
{Taylor} J.~H.,  {Stinebring} D.~R.,  1986, \mndoi [\araa]
  {10.1146/annurev.aa.24.090186.001441}, \href
  {http://adsabs.harvard.edu/abs/1986ARA%26A..24..285T} {24, 285}

\bibitem[\protect\citeauthoryear{{Taylor} et~al.,}{{Taylor}
  et~al.}{2012}]{Taylor2012}
{Taylor} G.~B.,  et~al., 2012, \mndoi [Journal of Astronomical Instrumentation]
  {10.1142/S2251171712500043}, \href
  {http://adsabs.harvard.edu/abs/2012JAI.....150004T} {1, 1250004}

\bibitem[\protect\citeauthoryear{{Thornton} et~al.,}{{Thornton}
  et~al.}{2013}]{Thornton2013}
{Thornton} D.,  et~al., 2013, \mndoi [Science] {10.1126/science.1236789}, \href
  {http://adsabs.harvard.edu/abs/2013Sci...341...53T} {341, 53}

\bibitem[\protect\citeauthoryear{{Thorsett}}{{Thorsett}}{1991}]{Thorsett1991}
{Thorsett} S.~E.,  1991, \mndoi [\apj] {10.1086/170355}, \href
  {http://adsabs.harvard.edu/abs/1991ApJ...377..263T} {377, 263}

\bibitem[\protect\citeauthoryear{{Tingay} et~al.,}{{Tingay}
  et~al.}{2013}]{Tingay2013}
{Tingay} S.~J.,  et~al., 2013, \mndoi [\pasa] {10.1017/pasa.2012.007}, \href
  {http://adsabs.harvard.edu/abs/2013PASA...30....7T} {30, e007}

\bibitem[\protect\citeauthoryear{{Tremblay} et~al.,}{{Tremblay}
  et~al.}{2015}]{Tremblay2015}
{Tremblay} S.~E.,  et~al., 2015, \mndoi [\pasa] {10.1017/pasa.2015.6}, \href
  {http://adsabs.harvard.edu/abs/2015PASA...32....5T} {32, e005}

\bibitem[\protect\citeauthoryear{{Xilouris}, {Kramer}, {Jessner}, {Wielebinski}
   \& {Timofeev}}{{Xilouris} et~al.}{1996}]{Xilouris1996}
{Xilouris} K.~M.,  {Kramer} M.,  {Jessner} A.,  {Wielebinski} R.,   {Timofeev}
  M.,  1996, \aap, \href {http://adsabs.harvard.edu/abs/1996A%26A...309..481X}
  {309, 481}

\bibitem[\protect\citeauthoryear{{van Haarlem} et~al.,}{{van Haarlem}
  et~al.}{2013}]{LOFARproject2013}
{van Haarlem} M.~P.,  et~al., 2013, \mndoi [\aap]
  {10.1051/0004-6361/201220873}, \href
  {http://adsabs.harvard.edu/abs/2013A%26A...556A...2V} {556, A2}

\bibitem[\protect\citeauthoryear{{von Hoensbroech}, {Kijak}  \&
  {Krawczyk}}{{von Hoensbroech} et~al.}{1998}]{Hoensbroech1998}
{von Hoensbroech} A.,  {Kijak} J.,   {Krawczyk} A.,  1998, \aap, \href
  {http://adsabs.harvard.edu/abs/1998A%26A...334..571V} {334, 571}

\makeatother
\end{thebibliography}

\onecolumn
\begin{center}
\begin{landscape}
\begin{ThreePartTable}
\begin{longtable}{@{}rrrrrrrrrrrrr@{}}
\caption{Flux density and other parameters for the 50 catalogued pulsars successfully detected in MWA-VCS archival data.} \label{tab:01} \\
\hline\hline
  \multicolumn{1}{c}{PSR} &
  \multicolumn{1}{c}{Period\tnote{1}} &
  \multicolumn{1}{c}{DM\tnote{2}} &
  \multicolumn{1}{c}{DM$_\text{psrcat}$\tnote{3}} &
  \multicolumn{1}{c}{Gain \tnote{4}} &
  \multicolumn{1}{c}{T$_{\text{sys}}$\tnote{5}} &
  \multicolumn{1}{c}{S/N$_{\text{peak}}$} &
  \multicolumn{1}{c}{S$_{185}$\tnote{6}} &
  \multicolumn{1}{c}{S$_{200}$\tnote{7}} &
  \multicolumn{1}{c}{S$_{400}$\tnote{8}} &
  \multicolumn{1}{c}{Fold time} &
  \multicolumn{1}{c}{Obs ID} \\
  \multicolumn{1}{c}{} &
  \multicolumn{1}{c}{(ms)} &
  \multicolumn{1}{c}{(cm$^{-3}$\,pc)} &
  \multicolumn{1}{c}{(cm$^{-3}$\,pc)} &
  \multicolumn{1}{c}{(K/Jy)} &
  \multicolumn{1}{c}{(K)} &
  \multicolumn{1}{c}{} &
  \multicolumn{1}{c}{(mJy)} &
  \multicolumn{1}{c}{(mJy)} &
  \multicolumn{1}{c}{(mJy)} &
  \multicolumn{1}{c}{(s)} &
  \multicolumn{1}{c}{} \\
\hline
\endfirsthead

\hline
  \multicolumn{1}{c}{PSR} &
  \multicolumn{1}{c}{Period\tnote{1}} &
  \multicolumn{1}{c}{DM\tnote{2}} &
  \multicolumn{1}{c}{DM$_\text{psrcat}$\tnote{3}} &
  \multicolumn{1}{c}{Gain \tnote{4}} &
  \multicolumn{1}{c}{T$_{\text{sys}}$\tnote{5}} &
  \multicolumn{1}{c}{S/N$_{\text{peak}}$} &
  \multicolumn{1}{c}{S$_{185}$\tnote{6}} &
  \multicolumn{1}{c}{S$_{200}$\tnote{7}} &
  \multicolumn{1}{c}{S$_{400}$\tnote{8}} &
  \multicolumn{1}{c}{Fold time} &
  \multicolumn{1}{c}{Obs ID} \\
  \multicolumn{1}{c}{} &
  \multicolumn{1}{c}{(ms)} &
  \multicolumn{1}{c}{(cm$^{-3}$\,pc)} &
  \multicolumn{1}{c}{(cm$^{-3}$\,pc)} &
  \multicolumn{1}{c}{(K/Jy)} &
  \multicolumn{1}{c}{(K)} &
  \multicolumn{1}{c}{} &
  \multicolumn{1}{c}{(mJy)} &
  \multicolumn{1}{c}{(mJy)} &
  \multicolumn{1}{c}{(mJy)} &
  \multicolumn{1}{c}{(s)} &
  \multicolumn{1}{c}{} \\
\hline
\endhead

 \hline
 \multicolumn{13}{r}{{Continued on next page}} \\
\endfoot

\hline \hline
\endlastfoot
    J0034$-$0534  &  1.877 &  13.77 &  13.77 &  0.0135 &  201 &  15.26 &  401$\pm$14 &  65$\pm$11 &  17 &  2496 &   1137236608 \\
    J0034$-$0721  &  943.045 &  11.12 &  10.92 &  0.0143 &  201 &  24.71 &  237$\pm$5 &  292$\pm$14 &  52 &  2496 &   1137236608 \\
    J0418$-$4154  &  757.094 &  24.54 &  24.54 &  0.0133 &  149 &  11.36 &  40$\pm$3 &       &          &  3712 &   1127329112 \\
    J0437$-$4715  &  5.757 &  2.65 &  2.64 &  0.0225 &  165 &  45.14 &  507$\pm$6 &  834$\pm$9 &  550 &  2496 &   1123367368 \\
    J0534$+$2200  &  33.7 &  56.78 &  56.79 &  0.0081 &  243 &    \tnote{\textdagger}   &    \tnote{\textdagger}   &       &  550 &  1024 &   1099414416 \\
    J0630$-$2834  &  1244.385 &  34.24 &  34.42 &  0.0201 &  213 &  52.95 &  271$\pm$3 &  463$\pm$5 &  206 &  4416 &   1101491208 \\
    J0742$-$2822  &  166.764 &  73.75 &  73.78 &  0.0129 &  213 &  15.05 &  168$\pm$8 &       &  296 &  3136 &   1101491208 \\
    J0820$-$1350  &  1238.043 &  40.94 &  40.94 &  0.0074 &  232 &  9.34 &  167$\pm$13 &  160$\pm$7 &  102 &  1024 &   1101925816 \\
    J0823$+$0159  &  864.865 &  23.73 &  23.73 &  0.0127 &  176 &  5.03 &  35$\pm$5 &       &  30 &  2688 &   1139324488 \\
    J0835$-$4510  &  89.39 &  67.97 &  67.99 &  0.0193 &  313 &    \tnote{\textdagger}   &    \tnote{\textdagger}   &  7075$\pm$207 &  5000 &  1216 &   1139239952 \\
    J0837$+$0610  &  1273.788 &  12.86 &  12.86 &  0.0163 &  176 &  159.12 &  654$\pm$9 &  286$\pm$13 &  89 &  2688 &   1139324488 \\
    J0837$-$4135  &  751.619 &  147.45 &  147.29 &  0.0233 &  313 &  16.29 &  151$\pm$4 &  95$\pm$16 &  197 &  1216 &   1139239952 \\
    J0855$-$3331  &  1267.621 &  86.64 &  86.64 &  0.0192 &  313 &  3.72 &  46$\pm$7 &  47$\pm$8 &  7.7 &  1216 &   1139239952 \\
    J0922$+$0638  &  430.631 &  27.3 &  27.3 &  0.0064 &  176 &  13.88 &  270$\pm$62 &  100$\pm$13 &  52 &  1344 &   1139324488 \\
    J0953$+$0755  &  253.09 &  2.96 &  2.97 &  0.0101 &  173 &  137.52 &  2123$\pm$16 &  1072$\pm$17 &  400 &  1216 &   1115381072 \\
    J1022$+$1001  &  16.456 &  10.25 &  10.25 &  0.0094 &  173 &  5.91 &  42$\pm$6 &       &  20 &  4864 &   1115381072 \\
    J1112$-$6926  &  820.447 &  148.4 &  148.4 &  0.0112 &  244 &  4.11 &  37$\pm$10 &       &  13 &  2496 &   1140972392 \\
    J1116$-$4122  &  943.199 &  40.53 &  40.53 &  0.0115 &  211 &  18.24 &  139$\pm$6 &  52$\pm$7 &  26 &  1984 &   1145367872 \\
    J1136$+$1551  &  1188.001 &  5.02 &  4.85 &  0.0029 &  173 &  50.47 &  1689$\pm$29 &  684$\pm$61 &  257 &  1792 &   1115381072 \\
    J1141$-$6545  &  394.036 &  116.08 &  116.08 &  0.0102 &  244 &  5.6 &  78$\pm$12 &       &          &  2496 &   1140972392 \\
    J1430$-$6623  &  785.453 &  65.08 &  65.3 &  0.0132 &  374 &  8.89 &  214$\pm$24 &  190$\pm$28 &  130 &  576 &   1131415232 \\
    J1440$-$6344  &  459.612 &  124.2 &  124.2 &  0.0121 &  374 &  5.13 &  176$\pm$34 &       &  21 &  576 &   1131415232 \\
    J1453$-$6413  &  179.49 &  71.03 &  71.07 &  0.0117 &  374 &  43.13 &  1244$\pm$20 &  684$\pm$23 &  230 &  576 &   1131415232 \\
    J1456$-$6843  &  263.382 &  8.62 &  8.6 &  0.0124 &  374 &  26.27 &  878$\pm$28 &  738$\pm$21 &  350 &  576 &   1131415232 \\
    J1507$-$4352  &  286.78 &  49.13 &  48.7 &  0.0111 &  778 &  13.27 &  384$\pm$27 &       &  16 &  2240 &   1121173352 \\
    J1534$-$5334  &  1368.968 &  24.82 &  24.82 &  0.0175 &  778 &  55.77 &  705$\pm$22 &       &  70 &  4864 &   1121173352 \\
    J1544$-$5308  &  178.565 &  35.16 &  35.16 &  0.0178 &  778 &  7.99 &  102$\pm$15 &       &  23 &  4864 &   1121173352 \\
    J1607$-$0032  &  421.829 &  10.6 &  10.68 &  0.0081 &  422 &  5.29 &  270$\pm$59 &  137$\pm$15 &  54 &  384 &   1117899328 \\
    J1645$-$0317  &  387.683 &  35.76 &  35.76 &  0.0085 &  832 &  27.36 &  2224$\pm$107 &  774$\pm$18 &  393 &  576 &   1116090392 \\
    J1709$-$1640  &  653.037 &  24.9 &  24.89 &  0.0171 &  832 &  15.97 &  271$\pm$18 &       &  47 &  3136 &   1116090392 \\
    J1731$-$4744  &  829.969 &  123.33 &  123.33 &  0.022 &  832 &  29.15 &  424$\pm$8 &  325$\pm$28 &  190 &  3840 &   1091793216 \\
    J1751$-$4657  &  742.405 &  20.38 &  20.4 &  0.0205 &  832 &  29.96 &  411$\pm$10 &       &  70 &  3840 &   1091793216 \\
    J1752$-$2806  &  562.521 &  50.25 &  50.37 &  0.0199 &  1021 &  116.02 &  2441$\pm$28 &  1504$\pm$269 &  1100 &  2432 &   1107478712 \\
    J1820$-$0427  &  598.044 &  84.69 &  84.44 &  0.0061 &  832 &  7.47 &  1035$\pm$120 &  499$\pm$51 &  157 &  1792 &   1116090392 \\
    J1823$-$3106  &  284.069 &  50.37 &  50.24 &  0.016 &  567 &  5.1 &  105$\pm$15 &       &  36 &  2944 &   1133329792 \\
    J1825$-$0935  &  768.973 &  19.38 &  19.38 &  0.0088 &  832 &  15.31 &  612$\pm$40 &       &  36 &  2112 &   1116090392 \\
    J1900$-$2600  &  612.241 &  37.99 &  37.99 &  0.0243 &  567 &  39.67 &  393$\pm$11 &  299$\pm$13 &  131 &  3712 &   1133329792 \\
    J1902$-$5105  &  1.742 &  36.24 &  36.25 &  0.0086 &  429 &  5.85 &  362$\pm$55 &       &          &  1664 &   1116787952 \\
    J1913$-$0440  &  826.027 &  89.61 &  89.39 &  0.0101 &  302 &  8.14 &  169$\pm$22 &  176$\pm$26 &  118 &  768 &   1097404000 \\
    J1917$+$1353  &  194.627 &  94.54 &  94.54 &  0.0106 &  385 &  8.28 &  237$\pm$30 &       &  43 &  1280 &   1148063920 \\
    J1921$+$2153  &  1337.39 &  12.46 &  12.44 &  0.0096 &  408 &  117.33 &  2112$\pm$49 &       &  57 &  2496 &   1095506112 \\
    J1932$+$1059  &  226.536 &  3.14 &  3.18 &  0.0111 &  408 &  20.74 &  362$\pm$15 &  501$\pm$47 &  303 &  3520 &   1095506112 \\
    J1935$+$1616  &  358.768 &  158.52 &  158.52 &  0.0115 &  408 &  6.82 &  106$\pm$9 &       &  242 &  3712 &   1095506112 \\
    J1943$-$1237  &  972.453 &  28.92 &  28.92 &  0.0136 &  265 &  7.16 &  68$\pm$8 &       &  12.9 &  1536 &   1152636328 \\
    J2018$+$2839  &  557.991 &  14.14 &  14.2 &  0.0077 &  331 &  11.71 &  561$\pm$41 &       &  314 &  576 &   1131957328 \\
    J2022$+$2854  &  343.427 &  24.63 &  24.63 &  0.0076 &  331 &  7.22 &  323$\pm$48 &       &  71 &  576 &   1131957328 \\
    J2046$-$0421  &  1547.1 &  36.23 &  35.8 &  0.0168 &  302 &  6.55 &  34$\pm$7 &       &  20 &  3456 &   1097404000 \\
    J2048$-$1616  &  1961.512 &  11.42 &  11.46 &  0.0199 &  265 &  17.4 &  77$\pm$5 &  169$\pm$8 &  116 &  4736 &   1152636328 \\
    J2145$-$0750  &  16.05 &  9 &  9 &  0.0149 &  259 &  27.57 &  282$\pm$20 &       &  100 &  4544 &   1118168248 \\
    J2241$-$5236  &  2.187 &  11.41 &  11.41 &  0.019 &  205 &  16.17 &  108$\pm$6 &  60$\pm$11 &          &  5056 &   1129464688 \\

\hline
\end{longtable}

    \begin{tablenotes}
    \item[1] Best period from our processing.
    \item[2] Best DM calculated from the data by maximising S/N in frequency vs. pulse phase. The effects of scattering and intrinsic profile evolution with frequency were not accounted for.
    \item[3] DM catalogued in PSRCAT.
    \item[4] As an aperture array, MWA's gain varies with pointing direction.
    \item[5] At our observing frequency, $T_\text{sys}$ is dominated by $T_\text{sky}$ which is, in turn, dominated by the synchrotron radiation from free electrons in the Galactic magnetic field.
    \item[6] Flux densities calculated from MWA-VCS detections. The quoted errors are the uncertainties from our estimation of S/N. They are from single epoch measurements and thus do not account for large flux density variations arisings from effects such as scintillation. By comparing the pulsars common to both this survey and \cite{Murphy2017}, the difference in flux density ranges from 1.5\% to 84.5\%.
    \item[7] Murphy et al. (2017) calculated 60 catalogued pulsars' flux densities from MWA continuum images.
    \item[8] Flux densities catalogued in PSRCAT.
    
    \item[\textdagger] We can not calculate the flux density for the Crab and Vela because of the significant scattering tails.
    \end{tablenotes}
\end{ThreePartTable}
\end{landscape}
\end{center}

\end{document}